\documentclass[preprint,3p,12pt]{elsarticle}
\usepackage{mathrsfs}
\usepackage{amsmath}
\usepackage{stmaryrd}
\usepackage{bbding}
\usepackage{dcolumn}
\usepackage{graphicx}
\usepackage{amsfonts}
\usepackage{amssymb}
\usepackage{psfrag}
\usepackage{wrapfig}
\usepackage{subfigure}
\usepackage{makeidx}
\usepackage{bm}
\usepackage{epsf}
\usepackage{epsfig}
\usepackage{setspace}
\usepackage{graphicx}
\usepackage{epstopdf}
\usepackage{psfrag}
\usepackage{subfigure}
\usepackage{color}
\usepackage{comment}

\newtheorem{rmk}{Remark}
\newtheorem{prop}{Proposition}

\begin{document}

\title{A Compact Fourth-order Gas-kinetic Scheme for the Euler and Navier-Stokes Solutions}

\author[HKUST1]{Xing Ji}
\ead{xjiad@connect.ust.hk}

\author[iapcm]{Liang Pan}
\ead{panliangjlu@sina.com}

\author[HKUST2]{Wei Shyy}
\ead{weishyy@ust.hk}

\author[HKUST1,HKUST2]{Kun Xu\corref{cor1}}
\ead{makxu@ust.hk}

\address[HKUST1]{Department of Mathematics, Hong Kong University of Science and Technology, Clear Water Bay, Kowloon, Hong Kong}
\address[HKUST2]{Department of Mechanical and Aerospace Engineering, Hong Kong University of Science and Technology, Clear Water Bay, Kowloon, Hong Kong}
\address[iapcm]{Institute of Applied Physics and Computational Mathematics, Beijing, 100088, China}
\cortext[cor1]{Corresponding author}

\begin{abstract}
In this paper, a fourth-order compact gas-kinetic scheme (GKS) is developed for the compressible Euler and
Navier-Stokes equations under the framework of two-stage
fourth-order temporal discretization and Hermite WENO (HWENO) reconstruction.
Due to the high-order gas evolution model, the GKS provides a time dependent gas distribution function
at a cell interface. This time evolution solution can be used not only for the
flux evaluation across a cell interface and its time derivative, but also time accurate evolution solution at a cell interface.
As a result, besides updating the conservative flow variables inside each control volume,
the GKS can get the cell averaged slopes inside each control volume as well
through the differences of flow variables at the cell interfaces.
So, with the updated flow variables and their slopes inside each cell, the HWENO reconstruction  can be naturally
implemented for the compact high-order reconstruction at the beginning of next step.
Therefore, a compact higher-order GKS, such as the
two-stages fourth-order compact scheme can be constructed.
This scheme is as robust as second-order one, but more accurate solution can be obtained.
In comparison with compact fourth-order DG method, the current scheme has only two stages instead of four within each time step
for the fourth-order temporal accuracy, and the CFL number used here can be on the order of $0.5$ instead of $0.11$ for the DG method.
Through this research, it concludes that the use of high-order time  evolution model rather than the first order Riemann solution
is extremely important for the design of robust, accurate, and efficient higher-order schemes for the compressible flows.

\end{abstract}

\begin{keyword}
two-stage fourth-order discretization,  compact gas-kinetic scheme, high-order evolution model,
Hermite WENO reconstruction.
\end{keyword}

\maketitle

\section{Introduction}
In past decades, there have been tremendous efforts on the
development of higher-order numerical methods for hyperbolic
conservation laws, and great success has been achieved. There are
many review papers and monographs about the current status of
higher-order schemes, which include essentially non-oscillatory
scheme (ENO) \cite{ENO1, ENO2, ENO3}, weighted essentially
non-oscillatory scheme (WENO) \cite{WENO-Liu, WENO-JS}, Hermite
weighted essentially non-oscillatory scheme (HWENO) \cite{HWENO1,
HWENO2,HWENO3}, and discontinuous Galerkin scheme (DG) \cite{DG2,
DG3}, etc.
For the WENO and DG methods,  two common ingredients are the use of Riemann solver for the
interface flux evaluation \cite{Riemann-appro} and the Runge-Kutta time-stepping for the high-order temporal accuracy
\cite{TVD-RK}. In terms of spatial accuracy, the WENO approach is based on large stencil and many cells are involved in the
reconstruction, which makes the scheme complicated in the application to complex geometry with unstructured mesh.
For the DG methods, the most attractive property is its compactness. Even with second order scheme stencil, higher-order spatial accuracy
can be achieved through the time evolution or direct update of higher-order spatial derivatives of flow variables.
However, in the flow simulations with strong shocks, the DG methods seem lack of robustness.
Great effort has been paid to limit the updated slopes or to find out the trouble cells beforehand.
Still, the development of WENO and DG methods is the main research direction for higher-order schemes.

In the above approaches, the first-order Riemann flux plays a key role for the flow evolution.
Recently, instead of Riemann solver,  many schemes have been developed based on
the time-dependent flux function, such as the generalized Riemann problem (GRP)
solver \cite{GRP1,GRP2,GRP3} and AEDR framework \cite{Riemann-appro}.
An outstanding method is the two-stage fourth order scheme for the Euler equations \cite{GRP-high},
where both the flux and time derivative of flux function are used in the construction of higher-order scheme.
A compact fourth order scheme can be also constructed under the GRP framework for the hyperbolic equations \cite{du-li}.
Certainly, the 4th-order 2-stages discretization has been used under other framework as well \cite{Seal2014,Seal2016}.

In the past years, the gas-kinetic scheme (GKS) has been developed
systematically \cite{GKS-Xu1,GKS-Xu2,UGKS}.
The flux evaluation in GKS is based on the kinetic model equation and its time evolution solution
from non-equilibrium towards to an equilibrium one.
In GKS, the spatial and temporal evolution of a gas distribution
function are fully coupled nonlinearly.
The comparison between  GRP and GKS has been presented in \cite{li-li-xu} and the main difference is that GKS
intrinsically provides a NS flux instead of inviscid one in GRP.
The third-order and fourth-order GKS can be developed as well without using Runge-Kutta time stepping technique, but their flux formulations
become  extremely complicated \cite{GKS-high2,liuna}, especially for multidimensional flow.
Under the framework of multiple stages and multiple derivatives (MSMD) technique for numerical solution of ordinary differential equations
\cite{hairer}, a two-stage fourth-order GKS with
second-order flux function was constructed for the Euler and Navier-Stokes equations
\cite{GKS-high3}. In comparison with the formal one-stage
time-stepping third-order gas-kinetic solver \cite{GKS-high1,
GKS-high2}, the fourth-order scheme not only reduces the complexity
of the flux function, but also improves the accuracy of the scheme,
although the third-order and fourth-order schemes take similar
computational cost. The robustness of the two-stage fourth-order GKS is as good as the second-order shock capturing scheme.
By combining the second-order or third-order GKS fluxes with the multi-stage
multi-derivative technique again, a family of high order gas-kinetic
methods has been constructed \cite{MSMD-GKS}. The above higher-order GKS uses the higher-order WENO reconstruction for spatial
accuracy. These schemes are not compact and have room for further improvement.

The GKS time dependent gas-distribution function
at a cell interface provides not only the
flux evaluation and its time derivative, but also time accurate flow variables
at a cell interface.
The design of compact GKS based on the cell averaged and cell interface values has been conducted before \cite{xu03,pan-xu1,pan-xu2}.
In the previous approach, the cell interface values are strictly enforced in the reconstruction, which may not be an appropriate approach.
In this paper, inspired by the Hermite WENO (HWENO) reconstruction and compact fourth order GRP scheme \cite{du-li},
instead of using the interface values we are going to get the slopes inside each control volume first,
then based on the cell averaged values and slopes inside each control volume the HWENO reconstruction is
implemented for the compact high-order reconstruction.
The higher-order compact GKS developed in this paper is basically a unified combination of three ingredients, which are the two-stage fourth-order framework for temporal discretization \cite{GKS-high3}, the higher-order gas evolution model for interface values and fluxes evaluations, and the HWENO reconstruction.
In comparison with the GRP based fourth-order compact scheme, the current GKS provides the time evolution
of cell interface values one order higher in time than that in the GRP formulation. This fact makes the GKS more flexible to be
extended to unstructured mesh, especially for the Navier-Stokes solutions.

The similarity and difference between the current compact 4th-order GKS and the 4th-order DG method include the followings.
Both schemes are time explicit, have the same order of accuracy, and use the identical compact stencil with the same HWENO reconstruction.
The standard Runge-Kutta DG scheme needs four stages within each time step to get a 4th-order temporal accuracy, and
the time step is on the order of CFL number $0.11$ from stability consideration.
For the 4th-order compact GKS, 2-stages are used for the same accuracy due to the use of both flux and its time derivative, and the time step
used in almost all calculations are on the order of CFL number $0.5$.
The updated slope in GKS comes from the explicit evolution solution of flow variables at the cell interfaces,
and the slope is obtained through the Gauss's theorem. The slope in DG method evolves through weak DG formulation.
The dynamic difference in slope update deviates the GKS and the DG method.
For the 4th-order compact GKS, the HWENO is fully implemented without using any additional trouble cell or limiting technique.
At end, the 4th-order compact GKS solves the NS equations naturally, it has the same robustness as the 2nd-order shock capturing scheme, and
it is much more efficient and robust than the same order DG method.

This paper is organized as follows. The brief review of the
gas-kinetic flux solver is presented in Section 2. In Section 3, the general
formulation for the two-stage temporal discretization is introduced.
In Section 4, the compact gas-kinetic
scheme with Hermite WENO reconstruction is given.
Section 5 includes inviscid and viscous test cases to validate
the current algorithm. The last section is the conclusion.

\section{Gas-kinetic evolution model}
The two-dimensional gas-kinetic BGK equation \cite{BGK-1} can be
written as
\begin{equation}\label{bgk}
f_t+\textbf{u}\cdot\nabla f=\frac{g-f}{\tau},
\end{equation}
where $f$ is the gas distribution function, $g$ is the corresponding
equilibrium state, and $\tau$ is the collision time. The collision
term satisfies the following compatibility condition
\begin{equation}\label{compatibility}
\int \frac{g-f}{\tau}\psi \text{d}\Xi=0,
\end{equation}
where $\psi=(1,u,v,\displaystyle \frac{1}{2}(u^2+v^2+\xi^2))$,
$\text{d}\Xi=\text{d}u\text{d}v\text{d}\xi_1...\text{d}\xi_{K}$, $K$
is the number of internal degree of freedom, i.e.
$K=(4-2\gamma)/(\gamma-1)$ for two-dimensional flows, and $\gamma$
is the specific heat ratio.

Based on the Chapman-Enskog expansion for BGK equation
\cite{GKS-Xu1}, the gas distribution function in the continuum
regime can be expanded as
\begin{align*}
f=g-\tau D_{\textbf{u}}g+\tau D_{\textbf{u}}(\tau
D_{\textbf{u}})g-\tau D_{\textbf{u}}[\tau D_{\textbf{u}}(\tau
D_{\textbf{u}})g]+...,
\end{align*}
where $D_{\textbf{u}}={\partial}/{\partial t}+\textbf{u}\cdot
\nabla$. By truncating on different orders of $\tau$, the
corresponding macroscopic equations can be derived. For the Euler
equations, the zeroth order truncation is taken, i.e. $f=g$. For the
Navier-Stokes equations, the first order truncated distribution function is
\begin{align}\label{ns}
f=g-\tau (ug_x+vg_y+g_t).
\end{align}
Based on the higher order truncations, the Burnett and super-Burnett
eqautions can be also derived \cite{GKS-Burnett,GKS-B}.

Taking moments of the BGK equation Eq.\eqref{bgk} and integrating with
respect to space, the semi-discrete finite volume scheme can be
written as
\begin{align}\label{finite}
\frac{\text{d}W_{ij}}{\text{d}t}=-\frac{1}{\Delta x}
(F_{i+1/2,j}(t)-F_{i-1/2,j}(t))-\frac{1}{\Delta y}
(G_{i,j+1/2}(t)-G_{i,j-1/2}(t)),
\end{align}
where $W_{ij}$ is the cell averaged value of conservative variables,
$F_{i+1/2,j}(t)$ and $G_{i,j+1/2}(t)$ are the time dependent
numerical fluxes at cell interfaces in $x$ and $y$ directions.
The Gaussian quadrature is used to achieve the accuracy in space, such that
\begin{align}\label{gauss}
F_{i+1/2,j}(t)=\frac{1}{\Delta
y}\int_{y_{j-1/2}}^{y_{j+1/2}}F_{i+1/2}(y,t)\text{d}y=\sum_{\ell=1}^2\omega_\ell F_{i+1/2,j_\ell}(t),
\end{align}
where $\omega_1=\omega_2=1/2$ are weights for the Gaussian
quadrature point $\displaystyle y_{j_\ell}=y_j+\frac{(-1)^{\ell-1}}{2\sqrt{3}}\Delta y$, $\ell= 1, 2$,
for a fourth-order accuracy.
$F_{i+1/2,j_\ell}(t)$ are numerical fluxes and can be obtained
as follows
\begin{align*}
F_{i+1/2,j_\ell}(t)=\int\psi u f(x_{i+1/2},y_\ell,t,u,v,\xi)\text{d}u\text{d}v\text{d}\xi,
\end{align*}
where $f(x_{i+1/2},y_\ell,t,u,v,\xi)$ is the gas distribution function at
the cell interface. In order to construct the numerical fluxes, the integral solution of
BGK equation Eq.\eqref{bgk} is used
\begin{equation}\label{integral1}
f(x_{i+1/2},y_\ell,t,u,v,\xi)=\frac{1}{\tau}\int_0^t g(x',y',t',u,v,\xi)e^{-(t-t')/\tau}dt'\\
+e^{-t/\tau}f_0(-ut,-vt,u,v,\xi),
\end{equation}
where $(x_{i+1/2}, y_\ell)=(0,0)$ is the location of cell interface,
and $x=x'+u(t-t')$ and $y=y'+v(t-t')$ are the trajectory of
particles. $f_0$ is the initial gas distribution function
representing the kinetic scale physics, $g$ is the corresponding
equilibrium state related to the hydrodynamic scale physics. The
flow behavior at cell interface depends on the ratio of time step
to the  local particle collision time $\Delta t/\tau$.


To construct time evolution solution of a gas distribution function at a cell interface,
the following notations are introduced first
\begin{align*}
a_1=&(\partial g/\partial x)/g, a_2=(\partial g/\partial y)/g,
A=(\partial g/\partial t)/g, B=(\partial A /\partial t),\\
d_{11}&=(\partial a_1/\partial x), d_{12}=(\partial a_1/\partial
y)=(\partial a_2/\partial x), d_{22}=(\partial a_2/\partial y),
\\
&b_{1}=(\partial a_1/\partial t)=(\partial A/\partial x),
b_{2}=(\partial a_2/\partial t)=(\partial A/\partial y),
\end{align*}
where $g$ is an equilibrium state, and the dependence on particle
velocity for each variable above, denoted as $\omega$, can be expanded as follows
\cite{GKS-Xu2}
\begin{align*}
\omega=\omega_{1}+\omega_{2}u+\omega_{3}v+\omega_{4}\displaystyle
\frac{1}{2}(u^2+v^2+\xi^2).
\end{align*}
For the kinetic part of the integral solution Eq.\eqref{integral1},
the initial gas distribution function can be constructed as
\begin{equation*}
f_0=f_0^l(x,y,u,v)\mathbb{H} (x)+f_0^r(x,y,u,v)(1- \mathbb{H}(x)),
\end{equation*}
where $\mathbb{H}(x)$ is the Heaviside function,  $f_0^l$ and $f_0^r$ are the
initial gas distribution functions on both sides of a cell
interface, which have one to one correspondence with the initially
reconstructed macroscopic variables. For the third-order scheme, the
Taylor expansion for the gas distribution function in space at
$(x,y)=(0,0)$ is expressed as
\begin{align}\label{third-1}
f_0^k(x,y)=f_G^k(0,0)&+\frac{\partial f_G^k}{\partial
x}x+\frac{\partial f_G^k}{\partial y}y+\frac{1}{2}\frac{\partial^2
f_G^k}{\partial x^2}x^2+\frac{\partial^2 f_G^k}{\partial x\partial
y}xy+\frac{1}{2}\frac{\partial^2 f_G^k}{\partial y^2}y^2,
\end{align}
where $k=l,r$. According to the Cpapman-Enskog expansion, $f_{G}^k$ can be written as
\begin{align}\label{third-2}
f_{G}^k=g_k-\tau(a_{1k}u+a_{2k}v+A_k)g_k,
\end{align}
where $g_l,g_r$ are the equilibrium states corresponding to the
reconstructed macroscopic variables $W_l, W_r$. Substituting
Eq.\eqref{third-1} and Eq.\eqref{third-2} into Eq.\eqref{integral1},
the kinetic part for the integral solution can be written as
\begin{equation}\label{dis1}
\begin{aligned}
&e^{-t/\tau}f_0^k(-ut,-vt,u,v,\xi) \\
=&C_7g_k[1-\tau(a_{1k}u+a_{2k}v+A_k)] \\
+&C_8g_k[a_{1k}u-\tau((a_{1k}^2+d_{11k})u^2+(a_{1k}a_{2k}+d_{12k})uv+(A_ka_{1k}+b_{1k})u)] \\
+&C_8g_k[a_{2k}v-\tau((a_{1k}a_{2k}+d_{12k})uv+(a_{2k}^2+d_{22k})v^2+(A_ka_{2k}+b_{2k})v)] \\
+&\frac{1}{2}C_9[g_k(a_{1k}^2+d_{11k})u^2+2(a_{1k}a_{2k}+d_{12k})uv+(a_{2k}^2+d_{22k})v^2],
\end{aligned}
\end{equation}
where the coefficients $a_{1k},...,A_k, k=l,r$ are defined according
to the expansion of $g_{k}$. After determining the kinetic part
$f_0$, the equilibrium state $g$ in the integral solution
Eq.\eqref{integral1} can be expanded in space and time as follows
\begin{align}\label{equli}
g=g_0+\frac{\partial g_0}{\partial x}x+&\frac{\partial g_0}{\partial
y}y+\frac{\partial g_0}{\partial t}t+\frac{1}{2}\frac{\partial^2
g_0}{\partial x^2}x^2+\frac{\partial^2 g_0}{\partial x\partial
y}xy+\frac{1}{2}\frac{\partial^2 g_0}{\partial
y^2}y^2\nonumber\\
&+\frac{1}{2}\frac{\partial^2 g_0}{\partial t^2}t^2+\frac{\partial^2
g_0}{\partial x\partial t}xt+\frac{\partial^2 g_0}{\partial
y\partial t}yt,
\end{align}
where $g_{0}$ is the equilibrium state located at interface, which
can be determined through the compatibility condition
Eq.\eqref{compatibility}
\begin{align}\label{compatibility2}
\int\psi g_{0}\text{d}\Xi=W_0=\int_{u>0}\psi
g_{l}\text{d}\Xi+\int_{u<0}\psi g_{r}\text{d}\Xi,
\end{align}
where $W_0$ are the macroscopic variables corresponding the
equilibrium state $g_{0}$. Substituting Eq.\eqref{equli} into Eq.\eqref{integral1}, the hydrodynamic part for the integral solution
can be written as
\begin{equation}\label{dis2}
\begin{aligned}
\frac{1}{\tau}\int_0^t
g&(x',y',t',u,v,\xi)e^{-(t-t')/\tau}dt' \\
=&C_1g_0+C_2g_0\overline{a}_1u+C_2g_0\overline{a}_2v+C_3g_0\overline{A}+\frac{1}{2}C_5g_0(\overline{A}^2+\overline{B}) \\
+&\frac{1}{2}C_4[g_0(\overline{a}_1^2+\overline{d}_{11})u^2+2(\overline{a}_1\overline{a}_2+\overline{d}_{12})uv+(\overline{a}_2^2+\overline{d}_{22})v^2] \\
+&C_6g_0[(\overline{A}\overline{a}_1+\overline{b}_{1})u+(\overline{A}\overline{a}_2+\overline{b}_{2})v],
\end{aligned}
\end{equation}
where the coefficients
$\overline{a}_1,\overline{a}_2,...,\overline{A},\overline{B}$ are
defined from the expansion of equilibrium state $g_0$. The
coefficients $C_i, i=1,...,9$ in Eq.\eqref{dis1} and Eq.\eqref{dis2}
are given by
\begin{align*}
C_1=1-&e^{-t/\tau}, C_2=(t+\tau)e^{-t/\tau}-\tau, C_3=t-\tau+\tau e^{-t/\tau},C_4=-(t^2+2t\tau)e^{-t/\tau},\\
C_5=&t^2-2t\tau,C_6=-t\tau(1+e^{-t/\tau}),C_7=e^{-t/\tau},C_8=-te^{-t/\tau},C_9=t^2e^{-t/\tau}.
\end{align*}
The coefficients in Eq.\eqref{dis1} and Eq.\eqref{dis2}
can be determined by the spatial derivatives of macroscopic flow
variables and the compatibility condition as follows
\begin{align}\label{co}
\begin{cases}
\displaystyle\langle a_1\rangle =\frac{\partial W }{\partial x},
\langle a_2\rangle =\frac{\partial W }{\partial y},
\langle A+a_1u+a_2v\rangle=0,\\
\displaystyle\langle a_1 ^2+d_{11}\rangle=\frac{\partial^2 W
}{\partial x^2}, \langle a_2 ^2+d_{22}\rangle=\frac{\partial^2 W
}{\partial y^2}, \displaystyle\langle
a_1a_2+d_{12}\rangle=\frac{\partial^2 W
}{\partial x\partial y}, \\
\displaystyle\langle(a_1^2+d_{11})u+(a_1a_2+d_{12})v+(Aa_1+b_1)\rangle=0,\\
\displaystyle\langle(a_2a_1+d_{21})u+(a_2^2+d_{22})v+(Aa_2+b_2)\rangle=0,\\
\displaystyle\langle(Aa_1+b_1)u+(Aa_2+b_2)v+(A^2+B)\rangle=0,
\end{cases}
\end{align}
where the superscripts or subscripts of these coefficients
$a_1,...,A, B$ are omitted for simplicity, more details about the
determination of coefficient can be found in  \cite{GKS-high1,GKS-high2}.
For the non-compact two stages fourth-order scheme \cite{GKS-high3}, theoretically
a second-order gas-kinetic solver is enough for accuracy requirement, where the above
third-order evolution solution reduces to the second-order one \cite{GKS-Xu2},
\begin{align}\label{flux}
f(x_{i+1/2,j_\ell},t,u,v,\xi)=&(1-e^{-t/\tau})g_0+((t+\tau)e^{-t/\tau}-\tau)(\overline{a}_1u+\overline{a}_2v)g_0\nonumber\\
+&(t-\tau+\tau e^{-t/\tau}){\bar{A}} g_0\nonumber\\
+&e^{-t/\tau}g_r[1-(\tau+t)(a_{1r}u+a_{2r}v)-\tau A_r)]H(u)\nonumber\\
+&e^{-t/\tau}g_l[1-(\tau+t)(a_{1l}u+a_{2l}v)-\tau A_l)](1-H(u)).
\end{align}

In this paper we will develop a compact scheme.
In order to make consistency between the flux evaluation and the interface value update,
a simplified 3rd-order gas distribution function  will be used \cite{GKS-zhou}.
With the introduction of the coefficients
\begin{equation}\label{flux_solver11}
\begin{split}
&a_x=a_1=g_x/g,a_y=a_2=g_y/g,a_t=A=g_t/g,
\\
&a_{xx}=g_{xx}/g,a_{xy}=g_{xy}/g,a_{yy}=g_{yy}/g,
\\
&a_{xt}=g_{xt}/g,a_{yt}=g_{yt}/g,a_{tt}=g_{tt}/g,
\end{split}
\end{equation}
the determination of these coefficients in Eq.({\ref{co}) is simplified to
\begin{align}\label{co2}
\begin{cases}
\displaystyle\left\langle a_x\right\rangle=\frac{\partial{W}}{\partial{x}},
\left\langle a_y\right\rangle=\frac{\partial{W}}{\partial{y}},
\left\langle a_t+a_xu+a_xv\right\rangle=0,
\\
\displaystyle\left\langle a_{xx}\right\rangle=\frac{\partial^2{W}}{\partial{x^2}},
\left\langle a_{xy}\right\rangle=\frac{\partial^2{W}}{\partial{xy}},
\left\langle a_{yy}\right\rangle=\frac{\partial^2{W}}{\partial{y^2}},
\\
\displaystyle\left\langle a_{xx}u+a_{xy}v+a_{xt}\right\rangle=0,
\\
\displaystyle\left\langle a_{xy}u+a_{yy}v+a_{yt}\right\rangle=0,
\\
\displaystyle\left\langle a_{xt}u+a_{yt}v+a_{tt}\right\rangle=0,
\end{cases}
\end{align}
and the final simplified distribution function from Eq.(\ref{integral1}), (\ref{dis1}) and (\ref{dis2}) becomes
\begin{equation}\label{3rd-simplify-flux}
\begin{split}
f(x_{i+1/2,j_\ell},y,t,u,v,\xi)=&g_0+\frac{1}{2}\bar{g}_{yy}y^2+\bar{g}_tt+\frac{1}{2}\bar{g}_{tt}t^2-\tau[(\bar{g}_t+u\bar{g}_x+v\bar{g}_y)+(\bar{g}_{tt}+u\bar{g}_{xt}+v\bar{g}_{yt})t]
\\&-e^{-t/\tau_n}[g_0-(u\bar{g}_x+v\bar{g}_y)t]
\\&+e^{-t/\tau_n}[g^l-(ug^l_x+vg_y^l)t]H(u)+e^{-t/\tau_n}[g^r-(ug^r_x+vg^r_y)t](1-H(u)).
\end{split}
\end{equation}
Both time-dependent interface flow variables and flux evaluations will be obtained from the
about Eq.(\ref{3rd-simplify-flux}).
With the same 3rd-order accuracy, the above simplified distribution function can speed up the flux calculation $4$ times
in comparison to the complete distribution function in 2-D case.

\section{Two-stage fourth-order temporal discretization}
Recently, the two-stage fourth-order temporal discretization was developed for the
generalized Riemann problem solver (GRP) \cite{GRP-high} and gas-kinetic scheme (GKS) \cite{GKS-high3}.
For conservation laws, the semi-discrete finite volume scheme is written as
\begin{align*}
\frac{\partial W_{ij}}{\partial t}=-\frac{1}{\Delta x}
(F_{i+1/2,j}(t)-F_{i-1/2,j}(t))-\frac{1}{\Delta y}
(G_{i,j+1/2}(t)-G_{i,j-1/2}(t)):=\mathcal{L}_{ij}(W),
\end{align*}
where $L_{ij}(W)$ is the numerical operator for spatial derivative of flux. With the following proposition, the two-stage
fourth-order scheme can be developed.
\begin{prop}
Consider the following time-dependent equation
\begin{align}\label{pde}
\frac{\partial W}{\partial t}=\mathcal {L}(W),
\end{align}
with the initial condition at $t_n$, i.e.,
\begin{align}\label{pde2}
W(t=t_n)=W^n,
\end{align}
where $\mathcal {L}$ is an operator for spatial derivative of flux. A
fourth-order temporal accurate solution for $W(t)$ at $t=t_n
+\Delta t$ can be provided by
\begin{equation}\label{step-hyper}
\begin{aligned}
&W^*=W^n+\frac{1}{2}\Delta t\mathcal
{L}(W^n)+\frac{1}{8}\Delta t^2\frac{\partial}{\partial
t}\mathcal{L}(W^n),\\
W^{n+1}&=W^n+\Delta t\mathcal
{L}(W^n)+\frac{1}{6}\Delta t^2\big(\frac{\partial}{\partial
t}\mathcal{L}(W^n)+2\frac{\partial}{\partial
t}\mathcal{L}(W^*)\big).
\end{aligned}
\end{equation}
where the time  derivatives are obtained by the Cauchy-Kovalevskaya
method
\begin{align*}
\frac{\partial W^n}{\partial t}=\mathcal{L}(W^n),&~\frac{\partial }{\partial t}\mathcal
{L}(W^n)=\frac{\partial }{\partial W}\mathcal
{L}(W^n)\mathcal {L}(W^n),\\
\frac{\partial W^*}{\partial t}=\mathcal{L}(W^*),&~\frac{\partial }{\partial t}\mathcal
{L}(W^*)=\frac{\partial }{\partial W}\mathcal
{L}(W^*)\mathcal {L}(W^*).
\end{align*}
The details of proof can be found in \cite{GRP-high}.
\end{prop}

In order to utilize the two-stage fourth-order temporal
discretization in the gas-kinetic scheme, the temporal derivatives
of the flux function need to be determined. While in order to obtain
the temporal derivatives at $t_n$ and $t_*=t_n + \Delta t/2$ with
the correct physics, the flux function should be approximated as a
linear function of time within the time interval. Let's first
introduce the following notation,
\begin{align*}
\mathbb{F}_{i+1/2,j}(W^n,\delta)
=\int_{t_n}^{t_n+\delta}F_{i+1/2,j}(W^n,t)dt&=\sum_{\ell=1}^2\omega_\ell \int_{t_n}^{t_n+\delta}\int
u \psi f(x_{i+1/2,j_\ell},t,u, v,\xi)\text{d}u\text{d}\xi\text{d}t,
\end{align*}
where the gas distribution function $f$ is provided in Eq.(\ref{3rd-simplify-flux}).
For convenience, assume $t_n=0$,
the flux in the time interval $[t_n, t_n+\Delta t]$ is expanded as
the following linear form
\begin{align}\label{expansion}
F_{i+1/2,j}(W^n,t)=F_{i+1/2,j}^n+ t \partial_t F_{i+1/2,j}^n  .
\end{align}
The coefficients $F_{i+1/2,j}^n$ and $\partial_tF_{i+1/2,j}^n$ can be
fully determined as follows
\begin{align*}
F_{i+1/2,j}(W^n,t_n)\Delta t&+\frac{1}{2}\partial_t
F_{i+1/2,j}(W^n,t_n)\Delta t^2 =\mathbb{F}_{i+1/2,j}(W^n,\Delta t) , \\
\frac{1}{2}F_{i+1/2,j}(W^n,t_n)\Delta t&+\frac{1}{8}\partial_t
F_{i+1/2,j}(W^n,t_n)\Delta t^2 =\mathbb{F}_{i+1/2,j}(W^n,\Delta t/2).
\end{align*}
By solving the linear system, we have
\begin{equation}\label{second}
\begin{aligned}
F_{i+1/2,j}(W^n,t_n)&=(4\mathbb{F}_{i+1/2,j}(W^n,\Delta t/2)-\mathbb{F}_{i+1/2,j}(W^n,\Delta t))/\Delta t,\\
\partial_t F_{i+1/2,j}(W^n,t_n)&=4(\mathbb{F}_{i+1/2,j}(W^n,\Delta t)-2\mathbb{F}_{i+1/2,j}(W^n,\Delta t/2))/\Delta
t^2.
\end{aligned}
\end{equation}
The flux for updating the intermediate state $W^*_{ij}$ is
\begin{align*}
\mathscr{F}_{i+1/2,j}^*&=\frac{1}{2}F_{i+1/2,j}(W^n,t_n)+\displaystyle\frac{\Delta
t}{8}\partial_t F_{i+1/2,j}(W^n,t_n),
\end{align*}
and the final expression of flux in unit time for updating the state $W^{n+1}_{ij}$ can be written as
\begin{align*}
\mathscr{F}_{i+1/2,j}^n&=F_{i+1/2,j}(W^n,t_n)+\displaystyle\frac{\Delta
t}{6}\big[\partial_t F_{i+1/2,j}(W^n,t_n)+2\partial_t F_{i+1/2,j}
(W^*,t_*)\big].
\end{align*}
Similarly, $(G_{i,j+1/2}(W^n,t_n)$ can be constructed
as well. In the fourth-order scheme, the first order time derivative of
the gas-distribution function is needed, which requires a higher order gas evolution than the Riemann problem.

Different from the Riemann problem with a constant state at a cell interface,
the gas-kinetic scheme provides a time evolution solution.
Taking moments of the time-dependent distribution function in Eq.(\ref{3rd-simplify-flux}), the
pointwise values at a cell interface can be obtained
\begin{align}\label{point}
W_{i+1/2,j_\ell}(t)=\int\psi f(x_{i+1/2,j_\ell},t,u,v,\xi)\text{d}u\text{d}v\text{d}\xi .
\end{align}
Similar to the
proposition for the two-stage temporal discretization, we have the
following proposition for the time dependent gas distribution
function at a cell interface
\begin{prop}
With the introduction of an intermediate state at $t_*=t_n+A\Delta t$,
\begin{align}\label{step-f1}
f^*=f^n+A\Delta tf_t^n+\frac{1}{2}A^2\Delta t^2f_{tt}^n,
\end{align}
the state $f^{n+1}$ is updated with the following formula
\begin{align}\label{step-f2}
f^{n+1}=f^n+\Delta t(B_0f_{t}^n+B_1f_{t}^*)+\frac{1}{2}\Delta
t^2\big(C_0f_{tt}^n+C_1f_{tt}^*\big),
\end{align}
and the solution $f^{n+1}$ at $(t=t_n +\Delta t)$ has fourth-order accuracy with
the following coefficients
\begin{align}\label{sol}
A=\frac{1}{2}, B_0=1, B_1=0, C_0=\frac13, C_1=\frac23.
\end{align}
\end{prop}
The proposition can be proved using the expansion
\begin{align*}
f^{n+1}=f^n+\Delta tf_{t}^n+\frac{\Delta
t^2}{2}f_{tt}^n+\frac{\Delta t^3}{6}f_{ttt}^n+\frac{\Delta
t^4}{24}f_{tttt}^n+\mathcal{O}(\Delta t^5).
\end{align*}
According to the definition of the intermediate state, the above expansion becomes
\begin{align*}
f^{n+1}-f^n&=\Delta t(B_0+B_1)f_t^n+\frac{\Delta
t^2}{2}(C_0+C_1+2B_1A)f_{tt}^n\nonumber\\
&+\frac{\Delta t^3}{2}(B_1A^2+C_1A )f_{ttt}^n+\frac{\Delta
t^4}{4}C_1A^2f_{tttt}^n+\mathcal{O}(\Delta t^5).
\end{align*}
To have a fourth-order accuracy for the interface value at $t^{n+1}$, the coefficients are uniquely
determined by Eq.\eqref{sol}.
Therefore, the macroscopic variables $W_{i+1/2}^{n+1}$ at a cell interface can be obtained by
taking moments of $f^{n+1}$ and the cell interface values can be used for the reconstruction at the beginning of next time step.

In order to utilize the two-stage fourth-order temporal
discretization for the gas distribution function, the third-order
gas-kinetic solver is needed. To construct the first and second
order derivative of the gas distribution function, the distribution function in Eq.(\ref{3rd-simplify-flux}) is approximated by the quadratic function
\begin{align*}
f(t)=f(x_{i+1/2,j_\ell},t,u, v,\xi)=f^n+
f_{t}^n(t-t^n)+\frac{1}{2}f_{tt}^n(t-t^n)^2.
\end{align*}
According to the gas-distribution function at
$t=0, \Delta t/2$, and $\Delta t$
\begin{align*}
f^n&=f(0),\\
f^n&+\frac{1}{2}f_{t}^n\Delta t+\frac{1}{8}f_{tt}^n\Delta t^2=f(\Delta t/2),\\
f^n&+f_{t}^n\Delta t+f_{tt}^n\Delta t^2=f(\Delta t),
\end{align*}
the coefficients $f^n, f_{t}^n$ and $f_{tt}^n$ can be
determined
\begin{align*}
f^n&=f(0),\\
f^n_t&=(4f(\Delta t/2)-3f(0)-f(\Delta t))/\Delta t,\\
f^n_{tt}&=4(f(\Delta t)+f(0)-2f(\Delta t/2))/\Delta t^2.
\end{align*}
Thus, $f^*$ and $f^{n+1}$ are fully determined at the cell interface.

\begin{rmk}

For smooth flow, the third-order evolution in Eq.(\ref{3rd-simplify-flux}) reduces to
\begin{align*}
f(x_{i+1/2,j_\ell},0,t,u,v,\xi)=&g_0+\bar{g}_tt+\frac{1}{2}\bar{g}_{tt}t^2
\end{align*}
The coefficients in Eq.(\ref{step-f1})and Eq.(\ref{step-f2}) can be
simplified as
\begin{align*}
f^n=g^n_0, f^n_t=\bar{g}^n_{t}, f^n_{tt}=\bar{g}^n_{tt},~\text{and}~
f^*=g^*_0, f^*_t=\bar{g}^*_{t}, f^*_{tt}=\bar{g}^*_{tt},
\end{align*}
and the cell interface values $W^{n+1}_{i+1/2}$ become
\begin{align*}
W^{n+1}_{i+1/2}=\int\psi f^{n+1}d\Xi,
\end{align*}
where  the gas distribution function $f^{n+1}$ has the form
\begin{align*}
f^{n+1}=&g^n_0+\Delta t \bar{g}^n_t+\frac{1}{6}\Delta
t^2(\bar{g}^n_{tt}+2\bar{g}^*_{tt}) .
\end{align*}
\end{rmk}

\begin{rmk}

For the scheme based on GRP in \cite{du-li}, the temporal evolution for the interface value is equivalent to
\begin{equation}\label{step-du}
\begin{split}
&f^*=f^n+\frac{1}{2}\Delta tf_t^n,\\
&f^{n+1}=f^n+\Delta tf_{t}^*,
\end{split}
\end{equation}
where a second-order evolution model is used at the cell interface.
This is the two step Runge-Kutta method with second order time accuracy for the updated cell interface values at time
step $t^{n+1}$. The order for the interface values is lower than that of GKS method. The method in \cite{du-li} may have
difficulty to get a compact 4th-order scheme in irregular mesh, such as unstructured one.

\end{rmk}	

\section{HWENO Reconstruction}
With the cell averaged values and pointwise values at cell interfaces, the
Hermite WENO (HWENO) reconstruction can be used for the
gas-kinetic scheme. The original HWENO
reconstruction \cite{HWENO1,HWENO2,HWENO3} was developed for the
hyperbolic conservation laws
\begin{align}\label{hyper11}
W_t+F(W)_x=0.
\end{align}
In order to construct the Hermite polynomials, the corresponding
equations for spatial derivative are used
\begin{align}\label{hyper2}
(W_x)_t+G(W,\Delta W)_x=0,
\end{align}
where $G(W,\Delta W)=F'(W)W_x=F'(W)(W_x)$. Therefore, in the previous HWENO approaches, the cell averaged values
$W_i$ and spatial derivative $(W_x)_i$ are updated by the following
semi-discrete finite volume scheme
\begin{align*}
\begin{cases}
\displaystyle\frac{\text{d}W_i}{\text{d}t}=-\frac{1}{\Delta x}(\widehat{F}_{i+1/2}-\widehat{F}_{i-1/2}),\\
\displaystyle\frac{\text{d}(W_x)_i}{\text{d}t}=-\frac{1}{\Delta
x}(\widehat{G}_{i+1/2}-\widehat{G}_{i-1/2}),
\end{cases}
\end{align*}
where $\widehat{F}_{i+1/2}$ and $\widehat{G}_{i+1/2}$ are the
corresponding numerical fluxes.
The above evolution solution for $W_x$ is different from the update of $W_x$ in the gas-kinetic scheme, which is shown in the following.

In the gas-kinetic scheme, the above equation for evolving the
spatial derivative $(W_x)_i$  is not needed, and the spatial
derivatives for all flow variables are updated  by the cell interface values with the help
of the Newton-Leibniz formula
\begin{align*}
(W_x)_{i}=\frac{1}{\Delta x}\int_{I_{i}}\frac{\partial
W}{\partial x}\text{d}x=\frac{1}{\Delta x}(W_{i+1/2}-W_{i-1/2}),
\end{align*}
where $W_{i+1/2}$ is provided by taking moments of the gas
distribution function at the cell interface according to
Eq.\eqref{step-f2}. With the cell averaged values $W_i$ and the cell
averaged spatial derivatives $(W_x)_{i}$, the HWENO method
can be directly applied for the spatial higher-order reconstruction.

\subsection{One-dimensional reconstruction}

Based on the cells $I_{i-1}, I_i, I_{i+1}$ and $I_{i+2}$, the compact HWENO reconstruction gives the
reconstructed variables $W_{i+1/2}^r$ and $W_{i+1/2}^l$
at both sides of the cell interface $x_{i+1/2}$ \cite{HWENO1}.
For cell $i$, with the reconstructed values of $W_{i-1/2}^r$, $W_{i+1/2}^l$ and cell averaged $W_i$, a parabolic distribution of
$W$ inside cell $i$ can be obtained, from which the initial condition for $f_0$ in cell $i$ are fully determined.
Then, the equilibrium state $g(0)$ at the cell interface is determined from  macroscopic variables $W(0)$
from the collision of left and right states  $W_{i+1/2}^r$ and $W_{i+1/2}^l$ according to
Eq.\eqref{compatibility2}.

To fully determine the slopes of the equilibrium state across the cell interface,
the conservative variables across the cell interface is expanded as
\begin{align*}
\overline{W}(x)=W_{0}+S_1(x-x_{i+1/2})+\frac{1}{2}S_2(x-x_{i+1/2})^2+\frac{1}{6}S_3(x-x_{i+1/2})^3+\frac{1}{24}S_4(x-x_{i+1/2})^4.
\end{align*}
With the following conditions,
\begin{align*}
\int_{I_{i+k}} \overline{W}(x)=W_{i+k}, k=-1,...,2,
\end{align*}
the derivatives are given by
\begin{align}\label{g0slope}
\begin{cases}
\displaystyle\overline{W}_x=S_1=\big[-\frac{1}{12}(W_{i+2}-W_{i-1})+\frac{5}{4}(W_{i+1}-W_{i})\big]/\Delta
x,\\
\displaystyle\overline{W}_{xx}=S_2=\big[-\frac{1}{8}(W_{i+2}+W_{i-1})+\frac{31}{8}(W_{i+1}+W_{i})-\frac{15}{2}W_0\big]/\Delta
x^2.
\end{cases}
\end{align}
Thus, the reconstruction for the initial data and the equilibrium
part are fully given in the one-dimensional case.

\subsection{Two dimensional reconstruction}

The direction by direction reconstruction strategy is applied on rectangular meshes \cite{accuracy-FVM}.
The HWENO reconstruction can be extended to 2-D straightforward.

Before introducing the reconstruction procedure, let's denote $\overline{W}$ as cell averaged, $\hat{W}$ as line averaged,
and $W$ as pointwise values. Here $W^{l,r}$ represent the reconstructed quantities on the left and right sides, which correspond to the non-equilibrium initial part in  GKS framework. Then, $W^e$ is the reconstructed equilibrium state.

At $t^n$ step, for cell $(i,j)$  the cell average quantities $\overline{W}^n_{i,j}$, $\overline{W}^n_{x,i,j}$, $\overline{W}^n_{y,i,j}$ are stored.
For a fourth order scheme, two Gaussian points in each interface are needed for numerical flux integration.
Our target is to construct $W^{l,r}, W^{l,r}_x, W^{l,r}_{xx}, W^{l,r}_{y}, W^{l,r}_{yy}, W^{l,r}_{xy}$ and $W^e, W^e_x, W^e_{xx}$, $W^e_{y}, W^e_{yy}, W^e_{xy}$ at each Gaussian point.
To obtain these quantities, four line averaged sloped $\hat{W}^n_{x,i,j_l}$, $\hat{W}^n_{y,i_l,j}$ are additionally evaluated, where $l=1,2$ represents the location Gaussian quadrature points in the corresponding direction.
For a better illustration, a schematic is plotted in Fig. \ref{hweno-stencil} and the reconstruction procedure for the Gaussian point $(i-1/2,j_1)$ is summarized as follows. Here the time level n is omitted.

\begin{figure}[!h]
\centering
\includegraphics[width=0.3\textwidth]{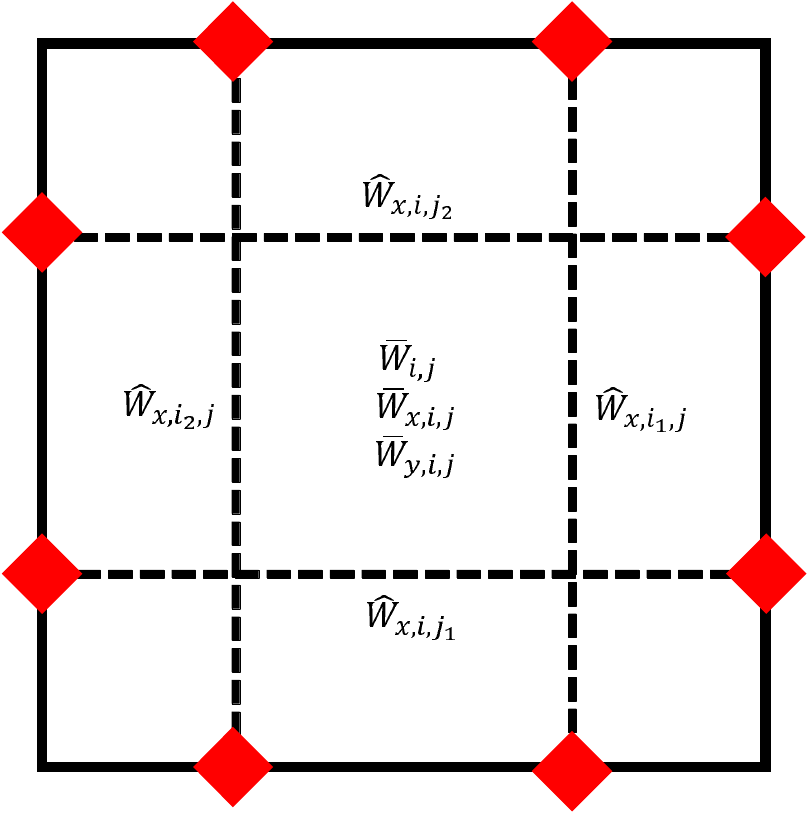}
\includegraphics[width=0.3\textwidth]{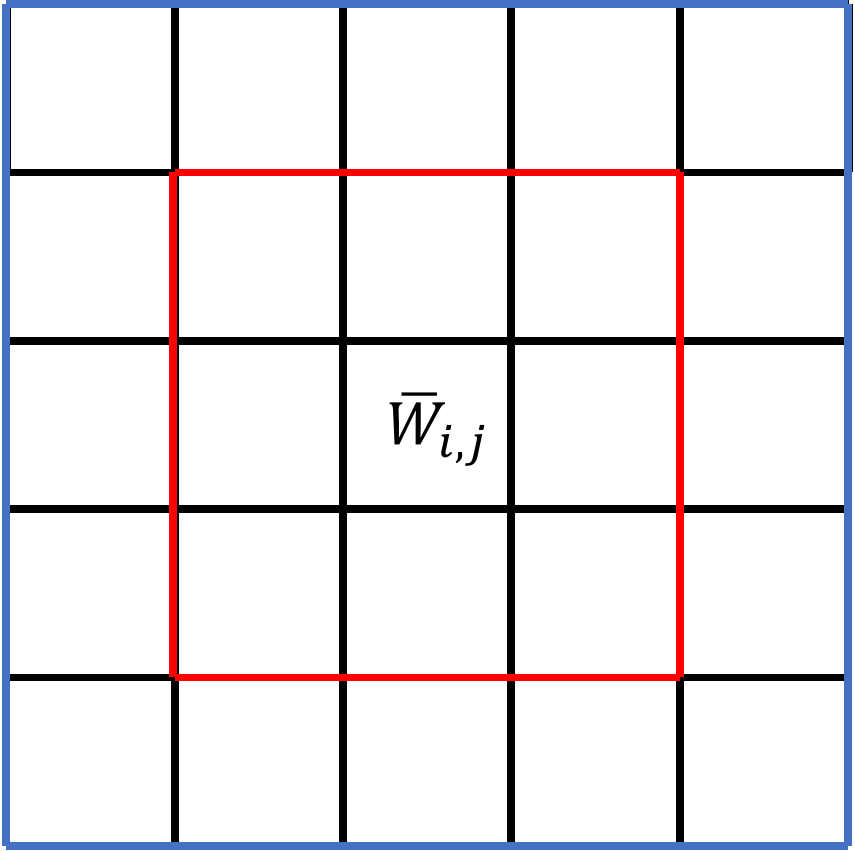}
\caption{\label{hweno-stencil}  The schematic for 2-D HWENO-GKS
reconstruction.
Left: Initial data for compact reconstruction at Gaussian points (in red color).
Right: Reconstruction stencils for current compact HWENO GKS (red box) and non-compact WENO GKS (blue box).  }
\end{figure}

\begin{enumerate}[Step 1.]
	\item To obtain the line average values, i.e. $\hat{W}_{i,j_l}$, we perform HWENO reconstruction in tangential direction by using   $\overline{W}_{i,j-1}, \overline{W}_{i,j}, \overline{W}_{i,j+1}$, and $\overline{W}_{y,i,j-1}, \overline{W}_{y,i,j+1}$. See Appendix 1 for details.
	\item With the reconstructed line average values i.e. $\hat{W}_{i-1,j_1}, \hat{W}_{i,j_1},\hat{W}_{i+1,j_1}$ and original $\hat{W}_{x,i-1,j_1},\hat{W}_{x,i+1,j_1}$, the one dimensional HWENO reconstruction is conducted and $W^{r}_{i-1/2,j_1},W^{l}_{i+1/2,j_1}$ are obtained.
	With the same derivative reconstruction method in \cite{GKS-high2}, $W^{r}_{x,i-1/2,j_1},W^{l}_{x,i+1/2,j_1}$ and $W^{r}_{xx,i-1/2,j_1},W^{l}_{xx,i+1/2,j_1}$ are constructed.
	\item To obtain the macro variables representing equilibrium state for each Gaussian point i.e. $\overline{W}_{i-1/2,j_1}$ with $\hat{W}^{l}_{x,i-1/2,j_1},\hat{W}^{r}_{x,i-1/2,j_1}$ by using Eq.(\ref{compatibility2}). ${W}^e_{x, i-1/2,j_1}$,$W^e_{xx, i-1/2,j_1}$ are calculated by Eq. (\ref{g0slope}).
	\item For the tangential derivatives, i.e.  $W^{r}_{y,i-1/2,j_1},W^{r}_{yy,i-1/2,j_1}$, a WENO-type reconstruction is adopted by using $W^{r}_{i-1/2,(j-1)_2},W^{r}_{i-1/2,j_1},W^{r}_{i-1/2,j_2},W^{r}_{i-1/2,(j+1)_1}$, see in Appendix 2. And $W^{r}_{xy,i-1/2,j_1}$ could be obtained in the same way with corresponding $W^{r}_x$.
	\item For the equilibrium state, a smooth third-order polynomial is constructed by  $W^e_{i-1/2,(j-1)_2},W^e_{i-1/2,j_1},W^e_{i-1/2,j_2},W^e_{i-1/2,(j+1)_1}$, and the tangential derivatives, i.e.  $W^e_{y,i-1/2,j_1},W^e_{yy,i-1/2,j_1}$ are also obtained. Then, $W^e_{xy,i-1/2,j_1}$ can be determined in the same way as for the corresponding
 $W^e_x$.
\end{enumerate}

Similar procedure can be performed to obtain all needed values at each Gaussian point.

After gas evolution process, the updated cell interface values are obtained, i.e. at time $t=t^*$ $W^*_{i \pm 1/2,j_l}$,$W^*_{i_l,j\pm 1/2}$,
as well as the the cell averaged slopes
\begin{align}
\label{updating-dell-avg}
\overline{W}^*_{x,i,j}&=\frac{1}{\Delta x} \sum_{l=1}^{2}(W^*_{i + 1/2,j_l}-W^*_{i - 1/2,j_l}),\\
\overline{W}^*_{y,i,j}&=\frac{1}{\Delta y}\sum_{l=1}^{2}(W^*_{i_l,j+1/2}-W^*_{i_l,j-1/2}),
\end{align}
according to the Gauss's theorem.
The cell averaged values are computed through conservation laws
\begin{align}
\label{updating-dell-avg}
\overline{W}^*_{i,j}=\overline{W}^n_{i,j}-\frac{1}{\Delta x} \sum_{l=1}^{2}(F^*_{i + 1/2,j_l}-F^*_{i - 1/2,j_l})
-\frac{1}{\Delta y}\sum_{l=1}^{2}(F^*_{i_l,j+1/2}-F^*_{i_l,j-1/2}),
\end{align}
where $F$ and $G$ are corresponding fluxes in x and y direction.
Lastly, in the rectangular case, the line averaged slopes are approximated by
\begin{align}
\label{updating-line-avg}
\hat{W}^*_{x,i,j_l}&=\frac{1}{\Delta x}(W^*_{i + 1/2,j_l}-W^*_{i - 1/2,j_l}),\\
\hat{W}^*_{y,i_l,j}&=\frac{1}{\Delta y}(W^*_{i_l,j+1/2}-W^*_{i_l,j-1/2}).
\end{align}

\section{Numerical examples}
In this section, numerical tests will be presented to validate the compact 4th-order GKS. For the
inviscid flow, the collision time $\tau$ is
\begin{align*}
\tau=\epsilon \Delta t+C\displaystyle|\frac{p_l-p_r}{p_l+p_r}|\Delta
t,
\end{align*}
where $\varepsilon=0.01$ and $C=1$. For the viscous flow, the collision time is related to the viscosity coefficient,
\begin{align*}
\tau=\frac{\nu}{p}+C \displaystyle|\frac{p_l-p_r}{p_l+p_r}|\Delta t,
\end{align*}
where $p_l$ and $p_r$ denote the pressure on the left and right
sides of the cell interface, $\nu$ is the dynamic viscous coefficient, and
$p$ is the pressure at the cell interface. In  smooth flow regions,
it will reduce to $\tau=\nu/p$. The ratio of specific heats takes
$\gamma=1.4$. The reason for including pressure jump term
in the particle collision time is to add artificial dissipation
in the discontinuous region, where the numerical cell size is not enough to resolve the shock structure,
and to keep the non-equilibrium in the kinetic formulation to mimic the real physical mechanism in the shock layer.

Same as many other higher-order schemes, all reconstructions will be done on the
characteristic variables. Denote $F(W)=(\rho U, \rho U^2+p, \rho
UV,U(\rho E+p))$ in the local coordinate. The Jacobian matrix
$\partial F/\partial W$ can be diagonalized by the right eigenmatrix
$R$. For a specific cell interface, $R_*$ is the right eigenmatrix
of $\partial F/\partial W^*$, and $W^*$ are the averaged
conservative flow variables from both sided of the cell interface. The characteristic variables
for reconstruction are defined as $U=R_*^{-1}W$.

The current compact 4th-order GKS is compared with the non-compact 4th-order WENO-GKS in \cite{GKS-high3,pan-li-xu}.
Both schemes take the same two Gaussian points at each cell interface in 2D case, and two stage fourth order time marching strategy for flux evaluation. The reconstruction is based on characteristic variables for both schemes and uses the same type non-linear weights of WENO-JS \cite{WENO-JS} in most cases. The main difference between them is on the initial data for reconstruction, where the large stencils used
in the normal WENO-GKS are  replaced by the local interface values in the compact HWENO-GKS.

\subsection{Accuracy tests}
The advection of density perturbation is tested, and the initial
condition is given as follows
\begin{align*}
\rho(x)=1+0.2\sin(\pi x),\ \  U(x)=1,\ \ \  p(x)=1, x\in[0,2].
\end{align*}
The periodic boundary condition is adopted, and the analytic
solution is
\begin{align*}
\rho(x,t)=1+0.2\sin(\pi(x-t)),\ \ \  U(x,t)=1,\ \ \  p(x,t)=1.
\end{align*}
In the computation, a uniform mesh with $N$ points is used. The time step $\Delta t=0.2\Delta x$ is fixed.
Before the full scheme using HWENO is tested, the order of accuracy for
the cell interface values will be validated firstly.
Here instead of using HWENO, we are going to use the cell interface values directly in the reconstruction for the compact GKS scheme.
Based on the compact stencil,
\begin{align*}
S=\{\overline{W}_{i-1},W_{i-1/2}, \overline{W}_{i}, W_{i+1/2},
\overline{W}_{i+1}\}
\end{align*}
with three cell averaged values and two cell interface values,
a fourth-order polynomial $W(x)$ can be constructed
according to the following constrains
\begin{align*}
\int_{I_{i+l}} W(x)\text{d}x=W_{i+l}, l=-1,0,1,~~~
W(x_{i+m-1/2})=W_{i+m-1/2}, m=0,1,
\end{align*}
where the cell interface value $W(x_{i+1/2})$ is equal to $W_{i+1/2}$ exactly.
Based on the above reconstruction, the compact scheme is expected to present a
fifth-order spatial accuracy and a fourth-order temporal
accuracy. The $L^1$ and $L^2$ errors and orders at
$t=2$ are given in Table.\ref{tab1}. This test shows that the cell interface updated values have the
the expected accuracy, which can be used in the spatial
reconstruction.
Nest, the full compact GKS is tested using the HWENO reconstruction, where the interface values are transferred into the cell
averaged slopes. For the HWENO compact GKS, the $L^1$ and $L^2$
errors and order of accuracy  at $t=2$ are shown in Table.\ref{tab2}. With
the mesh refinement, the expected order of accuracy is obtained as well.

\begin{table}[!h]
\begin{center}
\def\temptablewidth{0.85\textwidth}
{\rule{\temptablewidth}{0.5pt}}
\begin{tabular*}{\temptablewidth}{@{\extracolsep{\fill}}c|cc|cc}
mesh & $L^1$ error & convergence order ~ & $L^2$ error & convergence order  \\
\hline
10  & 1.2797E-003   &  ~~     &   9.8877E-004  &    ~     \\
20  & 7.2353E-005   &  4.1446 &   5.6650E-005  &    4.1254\\
40  & 3.3806E-006   &  4.4196 &   2.6547E-006  &    4.4154 \\
80  & 1.2863E-007   &  4.7159 &   1.0100E-007  &    4.7160\\
160 & 4.3188E-009   &  4.8965 &   3.3919E-009  &    4.8962\\
320 & 1.3819E-010   &  4.9658 &   1.0854E-010  &    4.9657\\
640 & 4.3517E-012   &  4.9890 &   3.4184E-012  &    4.9887
\end{tabular*}
{\rule{\temptablewidth}{0.5pt}}
\end{center}
\vspace{-4mm} \caption{\label{tab1} Advection of density
perturbation: accuracy test for the cell interface values in reconstructions.}
\begin{center}
\def\temptablewidth{0.85\textwidth}
{\rule{\temptablewidth}{0.5pt}}
\begin{tabular*}{\temptablewidth}{@{\extracolsep{\fill}}c|cc|cc}
mesh & $L^1$ error & convergence order ~ & $L^2$ error & convergence order  \\
\hline
10   & 2.666501e-04  &  ~      &  2.094924e-04  & ~      \\
20   & 1.082129e-05  &  4.6228 &  8.693374e-06  & 4.5908 \\
40   & 5.530320e-07  &  4.2904 &  4.967487e-07  & 4.1293 \\
80   & 3.251087e-08  &  4.0884 &  2.940079e-08  & 4.0786 \\
160  & 1.971503e-09  &  4.0436 &  1.769347e-09  & 4.0546 \\
320  & 1.210960e-10  &  4.0250 &  1.081183e-10  & 4.0325 \\
640  & 7.497834e-12  &  4.0135 &  6.675859e-12  & 4.0175
\end{tabular*}
{\rule{\temptablewidth}{0.5pt}}
\end{center}
\vspace{-4mm} \caption{\label{tab2} Advection of density
perturbation: accuracy test for HWENO compact GKS method at smooth
reconstruction.}
\end{table}

\begin{figure}[!h]
\centering
\includegraphics[width=0.485\textwidth]{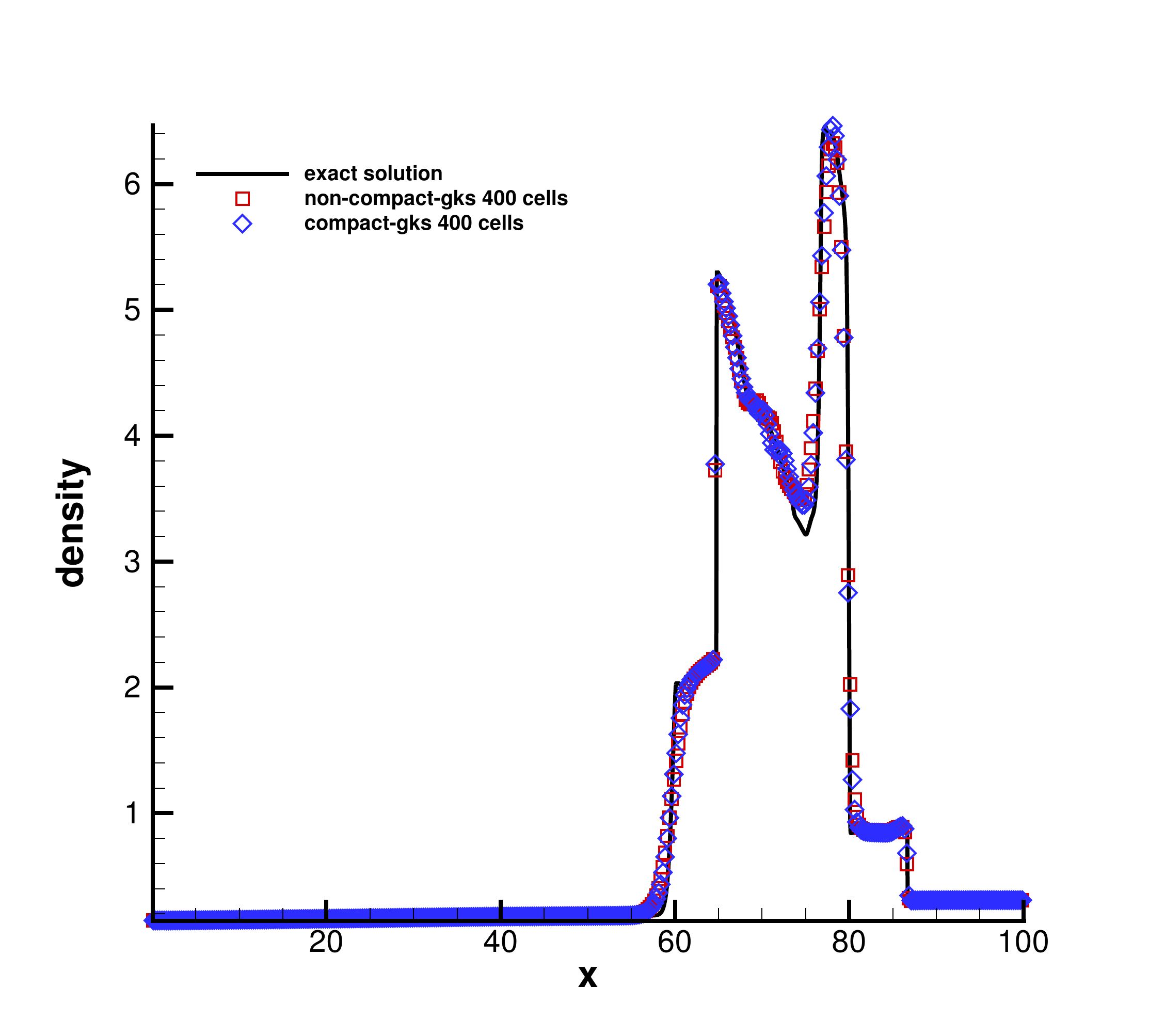}
\includegraphics[width=0.485\textwidth]{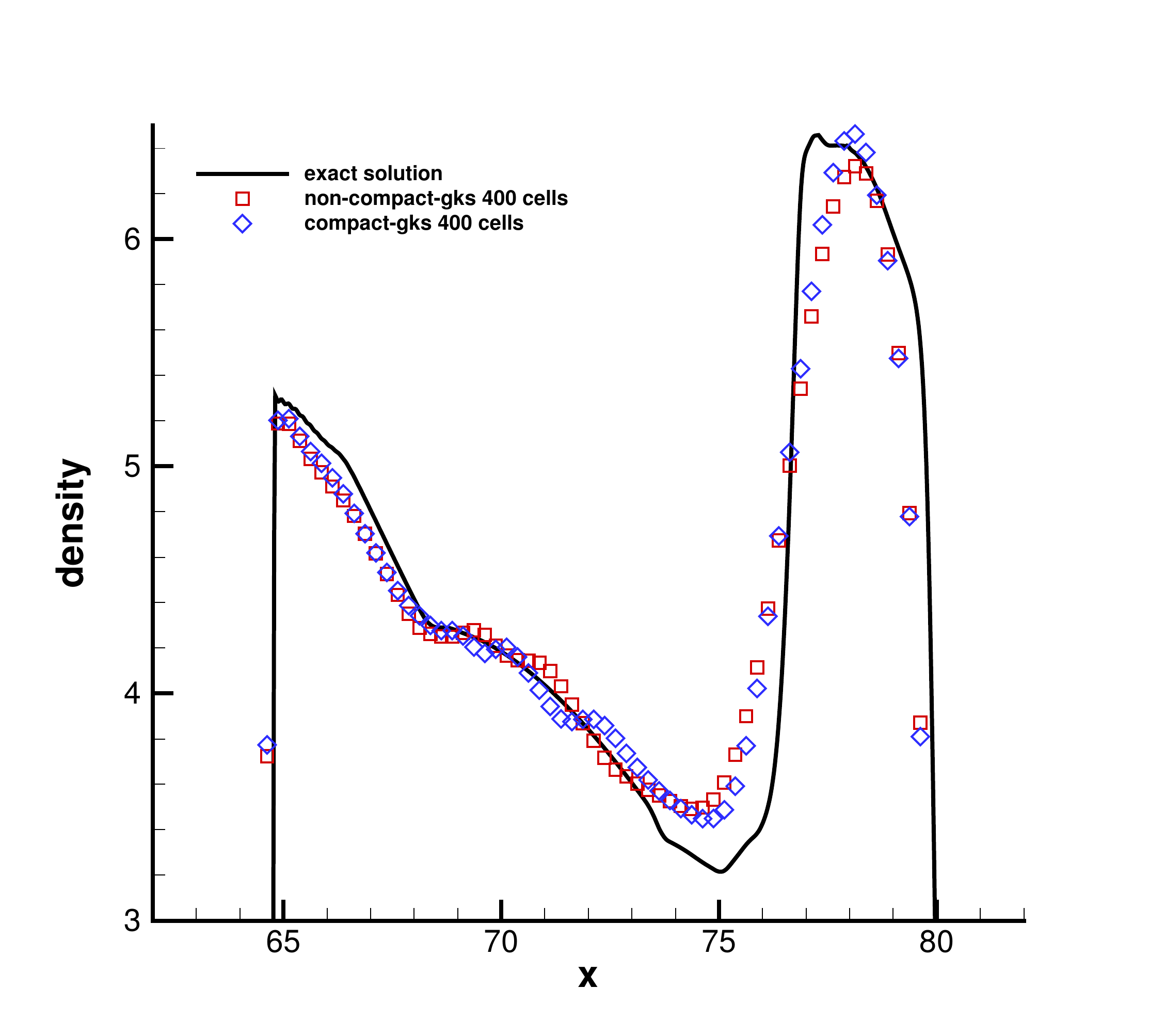}
\caption{\label{blastwave} Blast wave problem: the density
distribution and local enlargement at $t=3.8$ with $400$ cells.}
\end{figure}

\begin{figure}[!h]
\centering
\includegraphics[width=0.485\textwidth]{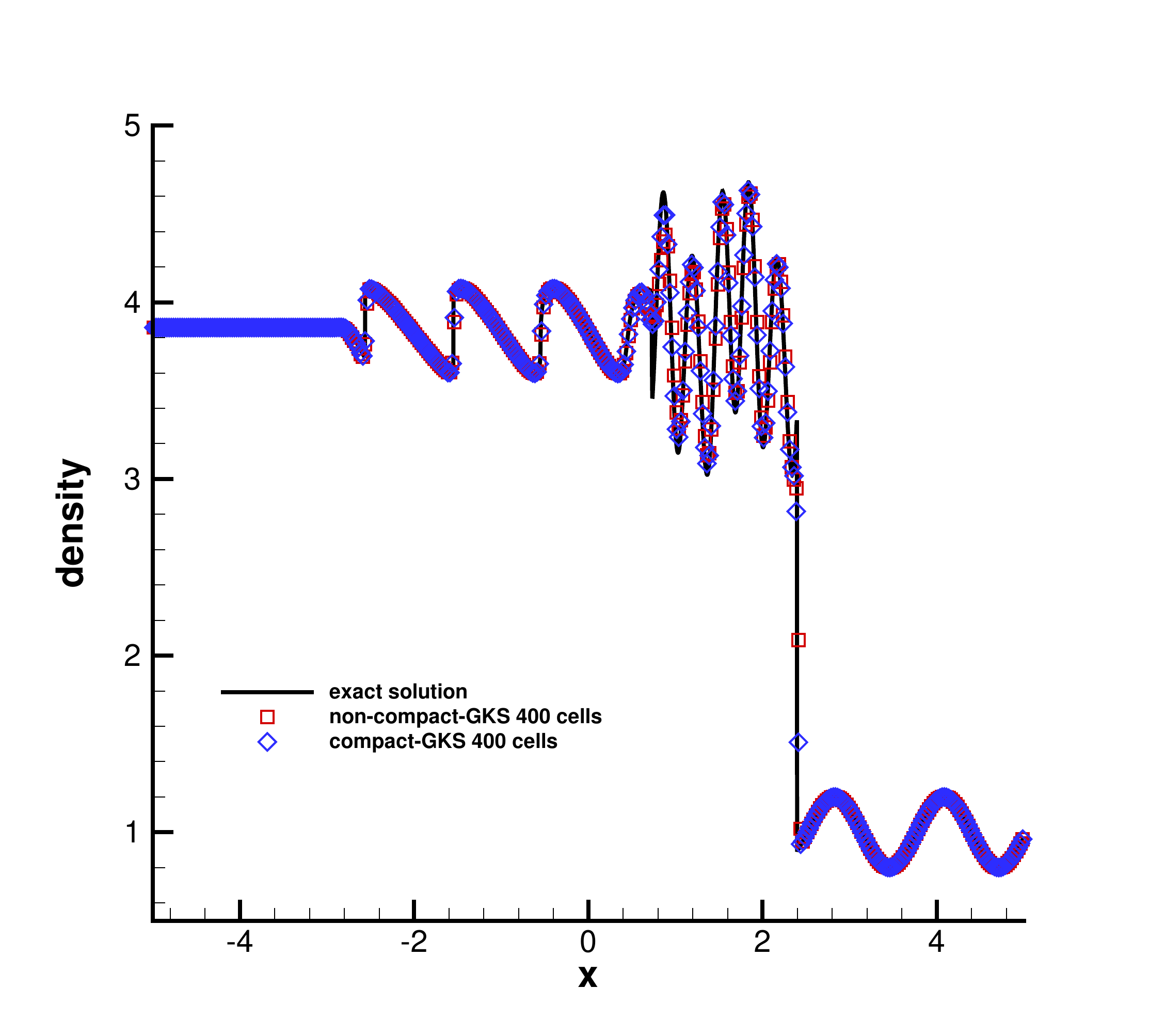}
\includegraphics[width=0.485\textwidth]{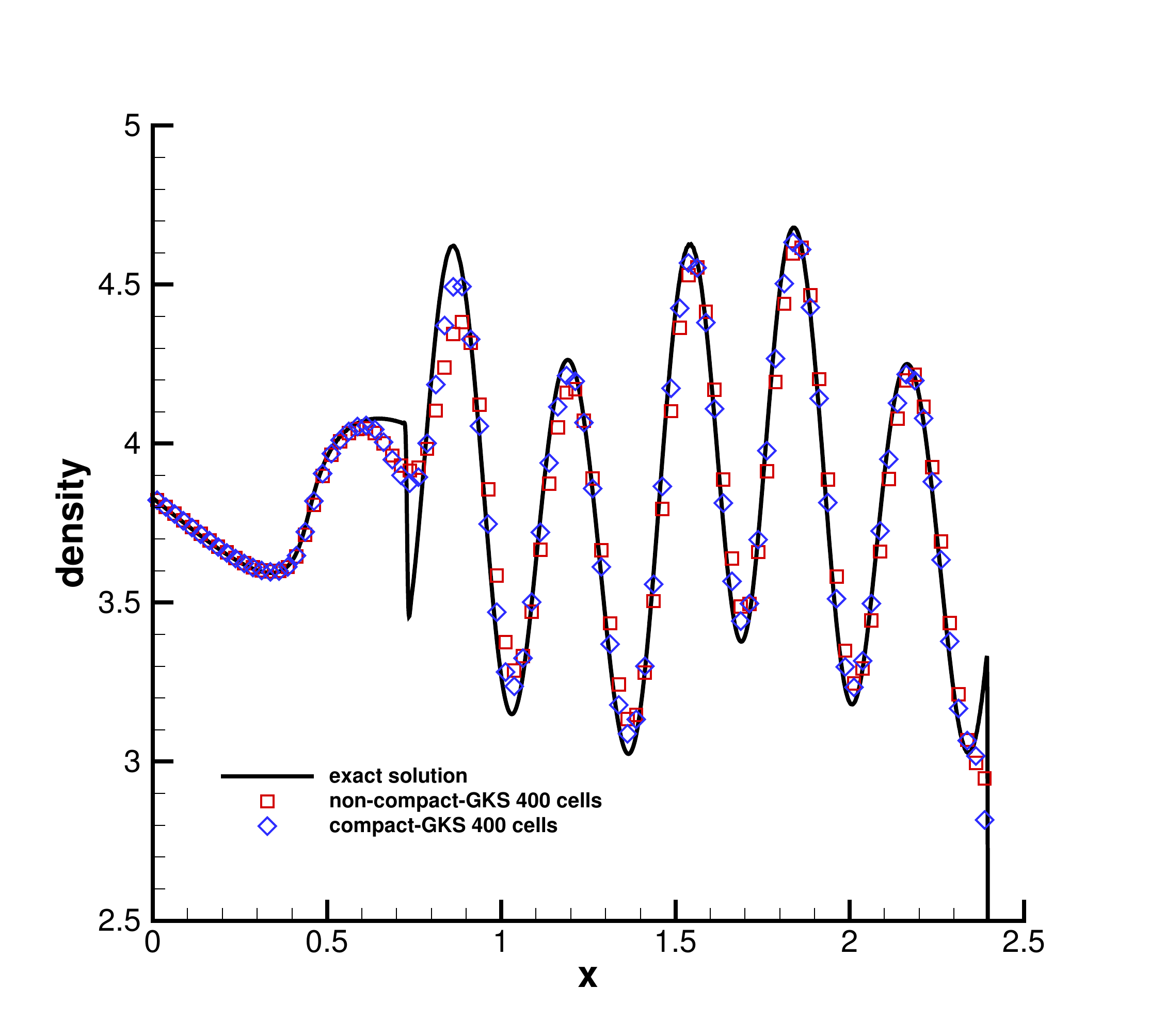}
\caption{\label{shuosher} Shu-Osher problem: the density
distributions and local enlargement at $t=1.8$ with $400$ cells.}
\centering
\includegraphics[width=0.485\textwidth]{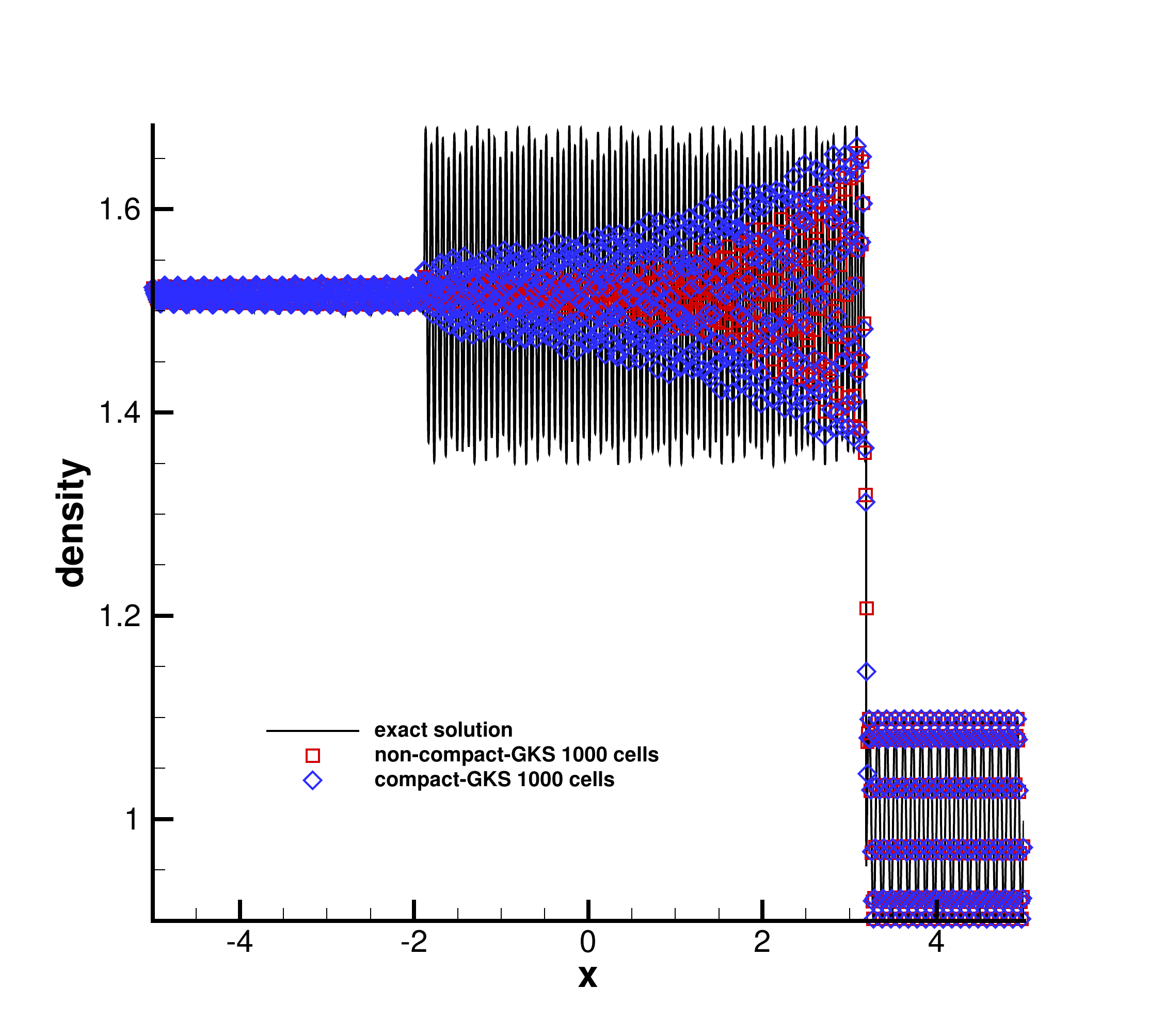}
\includegraphics[width=0.485\textwidth]{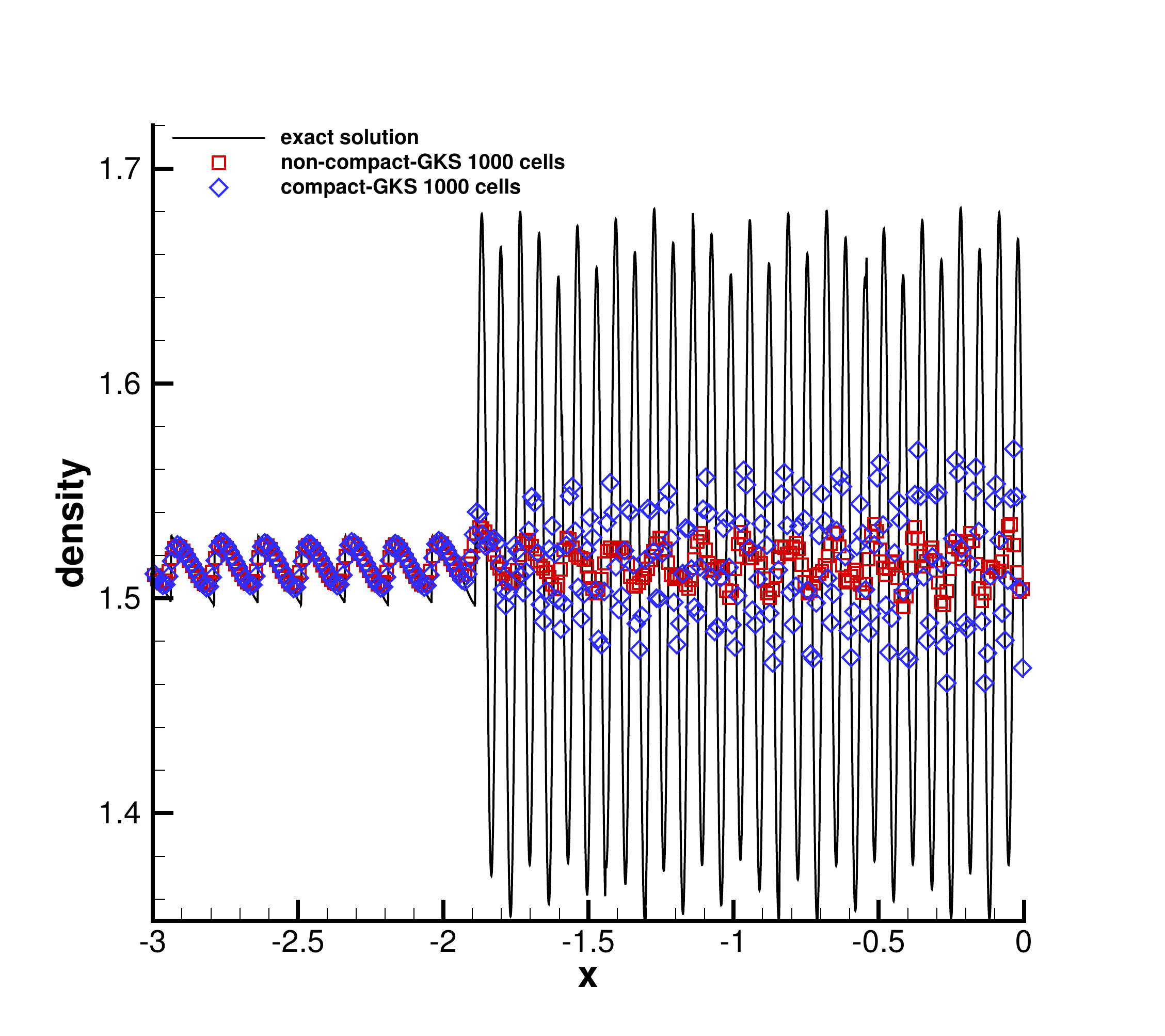}
\caption{\label{ttoro} Titarev-Toro problem: the density
distributions and local enlargement at $t=5$ with $1000$ cells.}
\end{figure}

\begin{figure}[!h]
	\centering
	\includegraphics[width=0.485\textwidth]{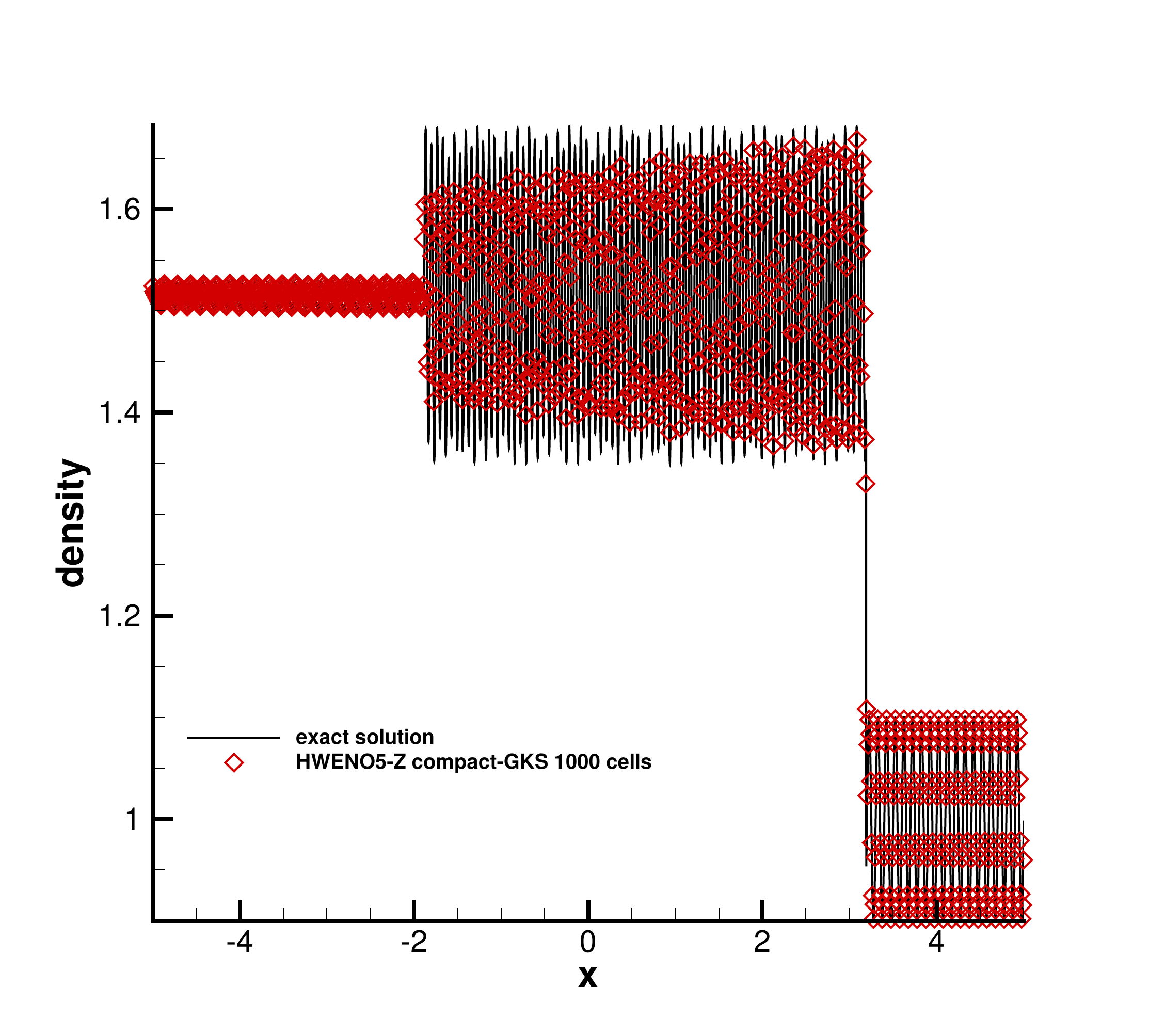}
	\includegraphics[width=0.485\textwidth]{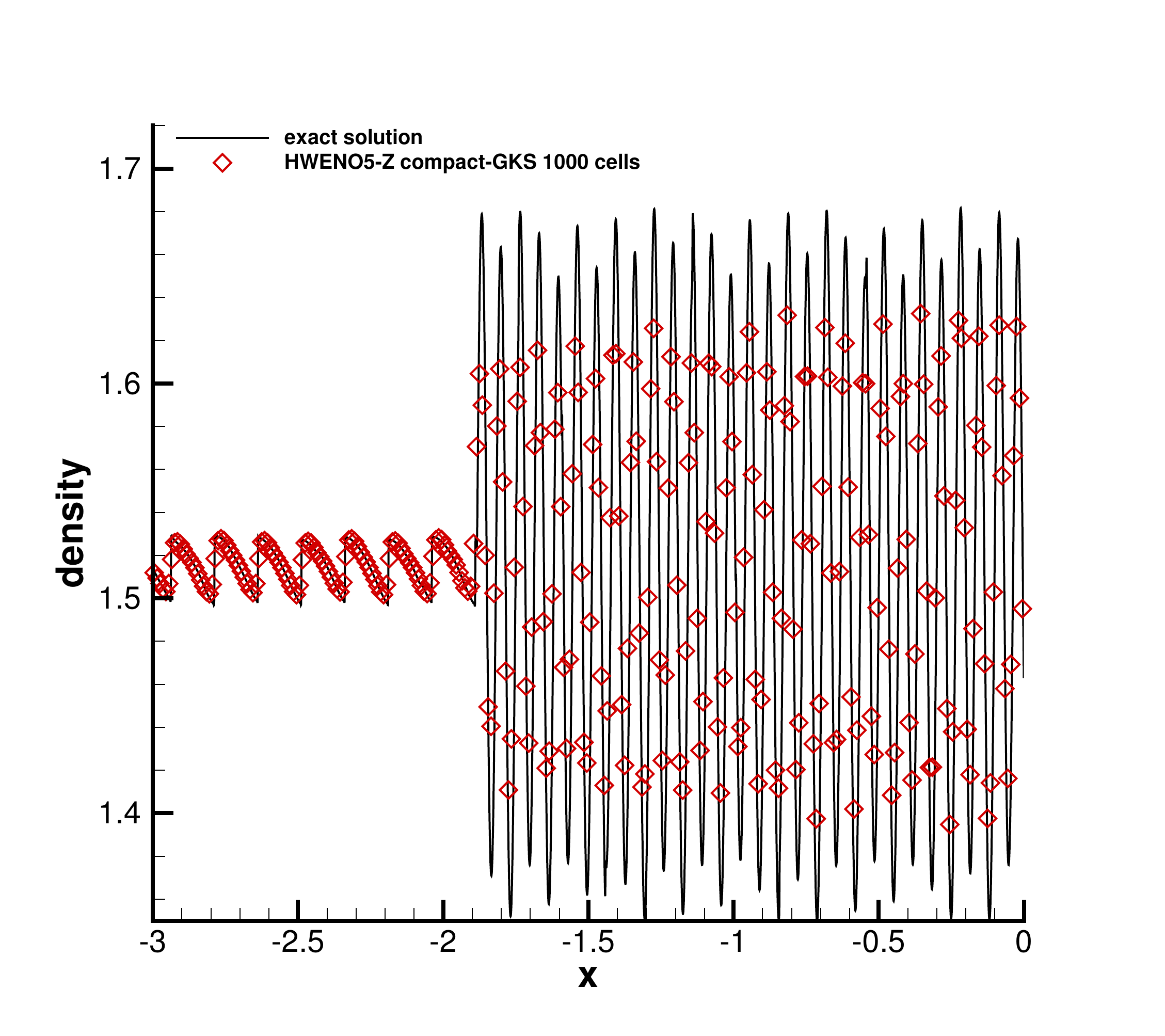}
	\caption{\label{ttoro2} Titarev-Toro problem:
		the results by using HWENO-Z reconstruction.}
\end{figure}

\subsection{One dimensional Riemann problems}
The first case is the Woodward-Colella blast wave problem \cite{Woodward-Colella}, and
the initial conditions are given as follows
\begin{align*}
(\rho,U,p) =\begin{cases}
(1, 0, 1000), & 0\leq x<10,\\
(1, 0, 0.01), & 10\leq x<90,\\
(1, 0, 100),  &  90\leq x\leq 100.
\end{cases}
\end{align*}
The computational domain is $[0,100]$, and the reflected boundary conditions are imposed on both ends.
The computed density profile and local enlargement with $400$ mesh
points and the exact solution at $t =3.8$ are shown in Fig.\ref{blastwave}.
The numerical results agree well with the exact solutions.
The scheme can resolve the wave profiles well,
particularly for the local extreme values.

The second one is the Shu-Osher problem \cite{ENO3}, and
the initial conditions are
\begin{align*}
(\rho,U,p) =\begin{cases}
(3.857134, 2.629369, 10.33333), &  -5<x \leq -4,\\
(1 + 0.2\sin (5x), 0, 1),  &  -4 <x<5.
\end{cases}
\end{align*}
As an extension of the Shu-Osher problem, the Titarev-Toro problem \cite{Titarev-Toro} is tested as well, and
the initial condition in this case is the following
\begin{align*}
(\rho,U,p) =\begin{cases}
(1.515695,0.523346,1.805),   & -5< x \leq -4.5,\\
(1 + 0.1\sin (20\pi x), 0, 1),  &  -4.5 <x <5.
\end{cases}
\end{align*}
In these two cases, the computational domain is $[-5, 5]$. The non-reflecting boundary condition is imposed on left end,
and the fixed wave profile is given on the right end.
Both compact GKS with HWENO and non-compact GKS with fifth-order WENO  are tested for these two cases.
The computed density profiles, local enlargements, and the exact solutions for the Shu-Osher problem with $400$ mesh
points at $t = 1.8$  and the Titarev-Toro problem with $1000$ mesh points at $t =5$  are shown in Fig.\ref{shuosher} and Fig.\ref{ttoro}, respectively.
Titarev-Toro problem is sensitive to reconstruction scheme \cite{WENO-Z,du-li}. Instead of WENO-JS used above for non-linear weights,
the WENO-Z weights can keep the same order of accuracy in extreme points.
Combing the HWENO-Z reconstruction and the compact GKS,
the result is shown in Fig.\ref{ttoro2}, which can be compared with the solution from the GRP method \cite{du-li}.

\subsection{Two-dimensional Riemann problems}
In the following, two examples of two-dimensional Riemann problems
are considered \cite{Case-lax}, which involve the interactions of
shocks and the interaction of contact continuities. The
computational domain  $[0,1]\times[0,1]$ is covered by  $500\times500$ uniform
mesh points, where non-reflecting boundary conditions are
used in all boundaries. The initial conditions for the first problem
are
\begin{equation*}
(\rho,U,V,p) =\left\{\begin{array}{ll}
         (1.5,0,0,1.5), \ \ \ &x>0.7,y>0.7,\\
         (0.5323,1.206,0,0.3), &x<0.7,y>0.7,\\
         (0.138,1.206,1.206,0.029), &x<0.7,y<0.7,\\
         (0.5323,0,1.206,0.3), &x>0.7,y<0.7.
                          \end{array} \right.
                          \end{equation*}
Four initial shock waves interact with each other and result in a
complicated flow pattern. The density distributions calculated by compact and non-compact GKS
with HWENO and WENO reconstructions are presented at $t=0.6$ in Fig.
\ref{rm2d-shock-500-1}. From the analysis in \cite{Case-lax}, the
initial shock wave $S^-_{23}$ bifurcates at the trip point into a
reflected shock wave, a Mach stem, and a slip line. The reflecting
shock wave interacts with the shock wave $S^-_{12}$ to produce a new
shock. The small scale flow structures are well resolved by the
current scheme.

\begin{figure}[!h]
\centering
\includegraphics[width=0.48\textwidth]{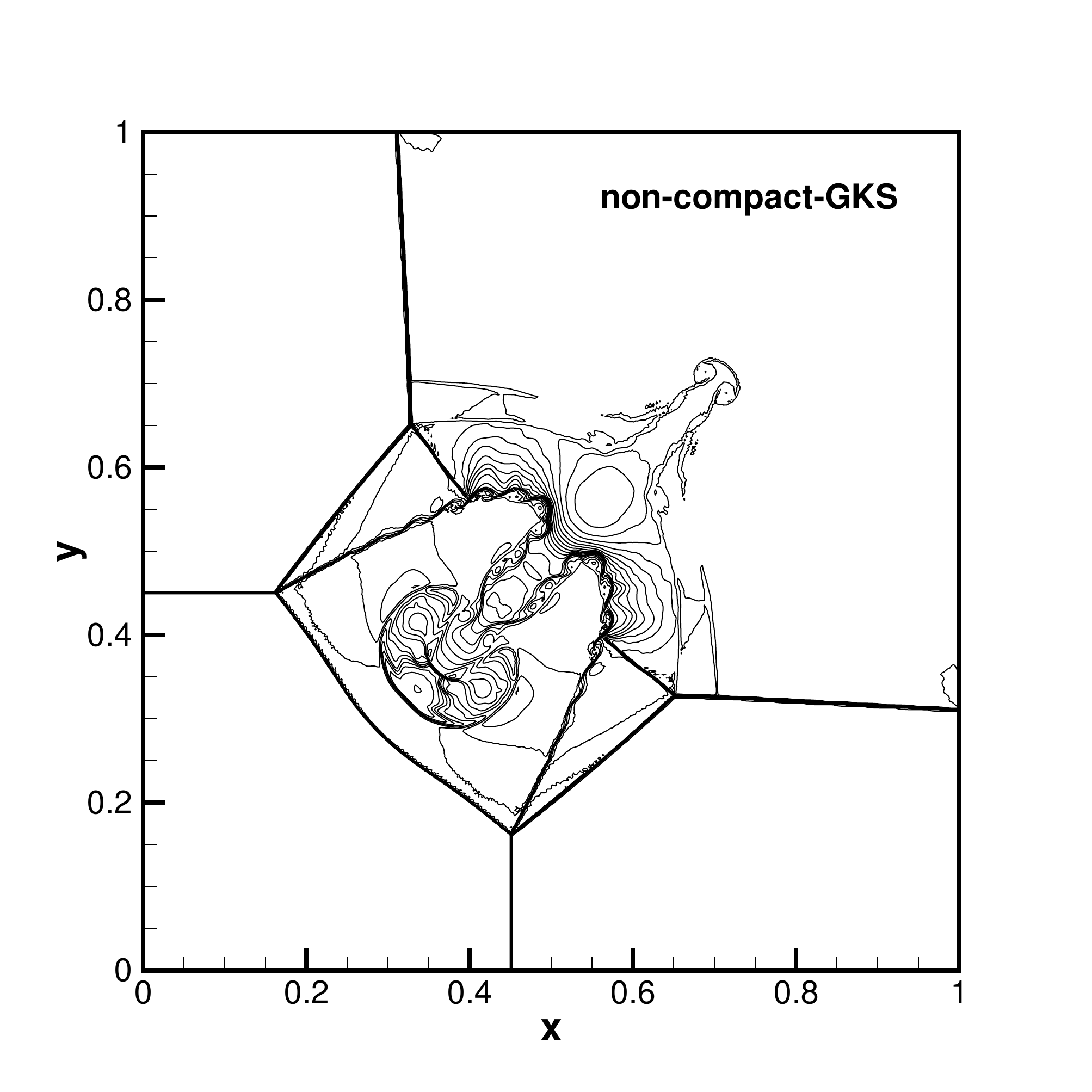}
\includegraphics[width=0.48\textwidth]{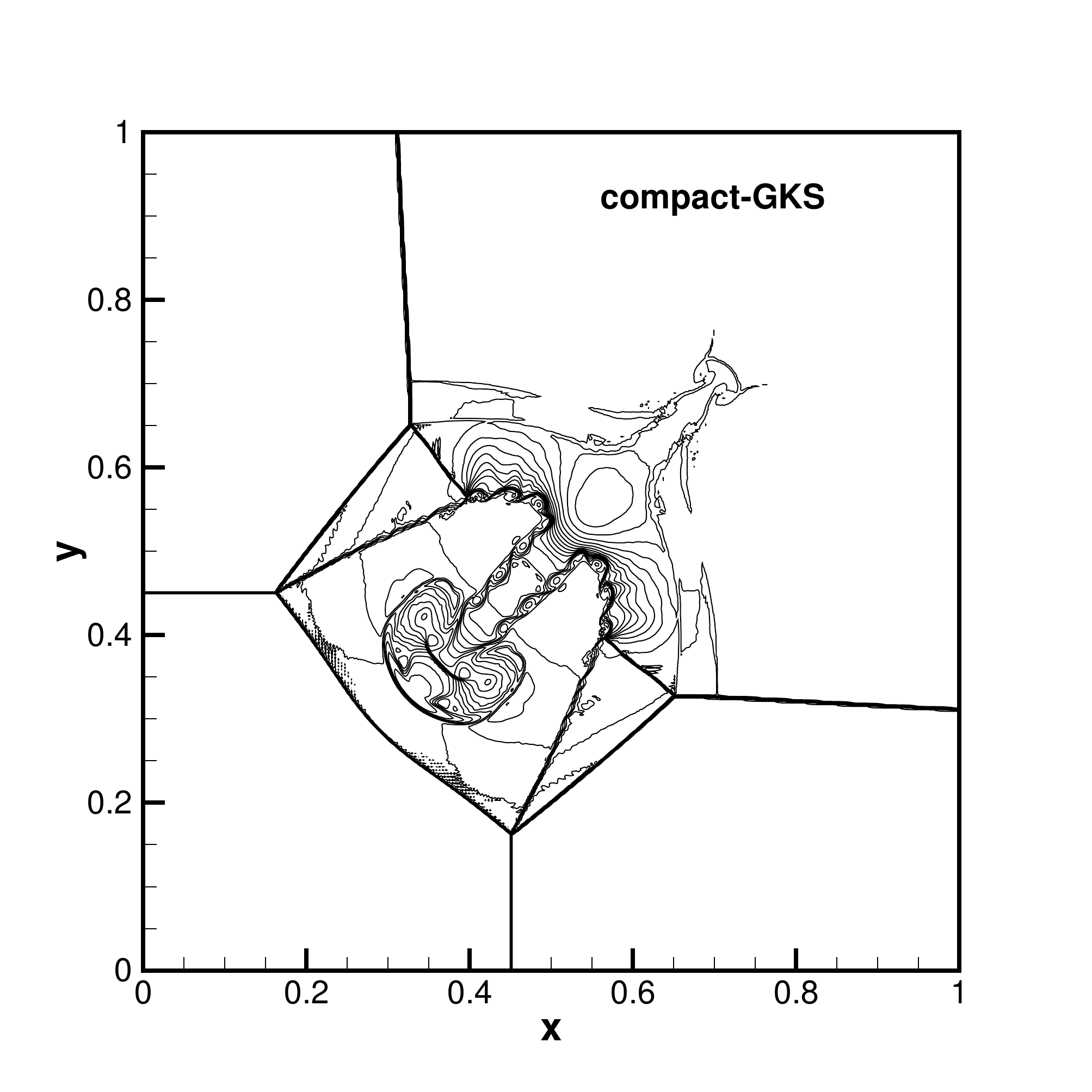}
\caption{\label{rm2d-shock-500-1} Two-dimensional Riemann problem:
the density distribution of four shock-interaction at $t=0.6$.}
\centering
\includegraphics[width=0.48\textwidth]{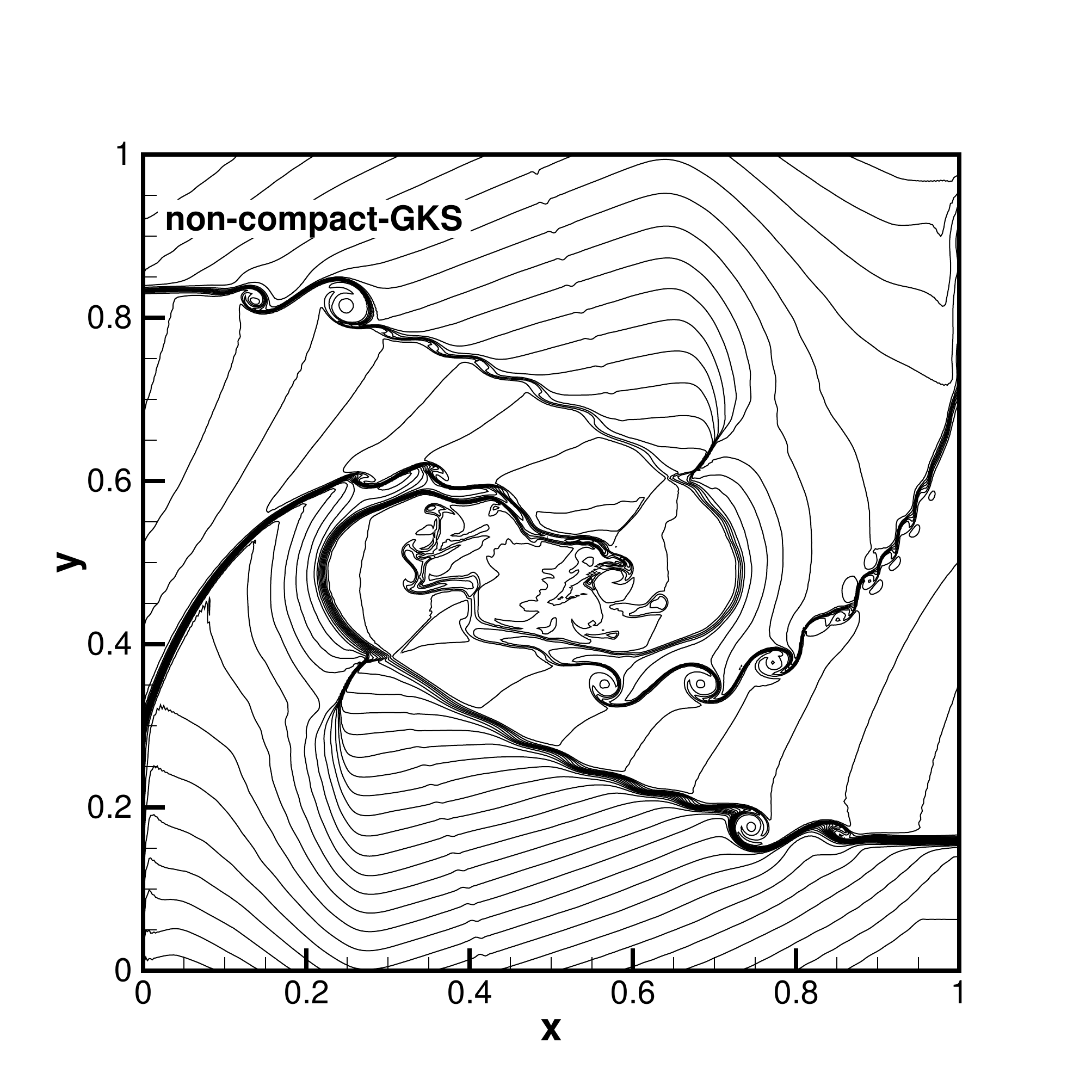}
\includegraphics[width=0.48\textwidth]{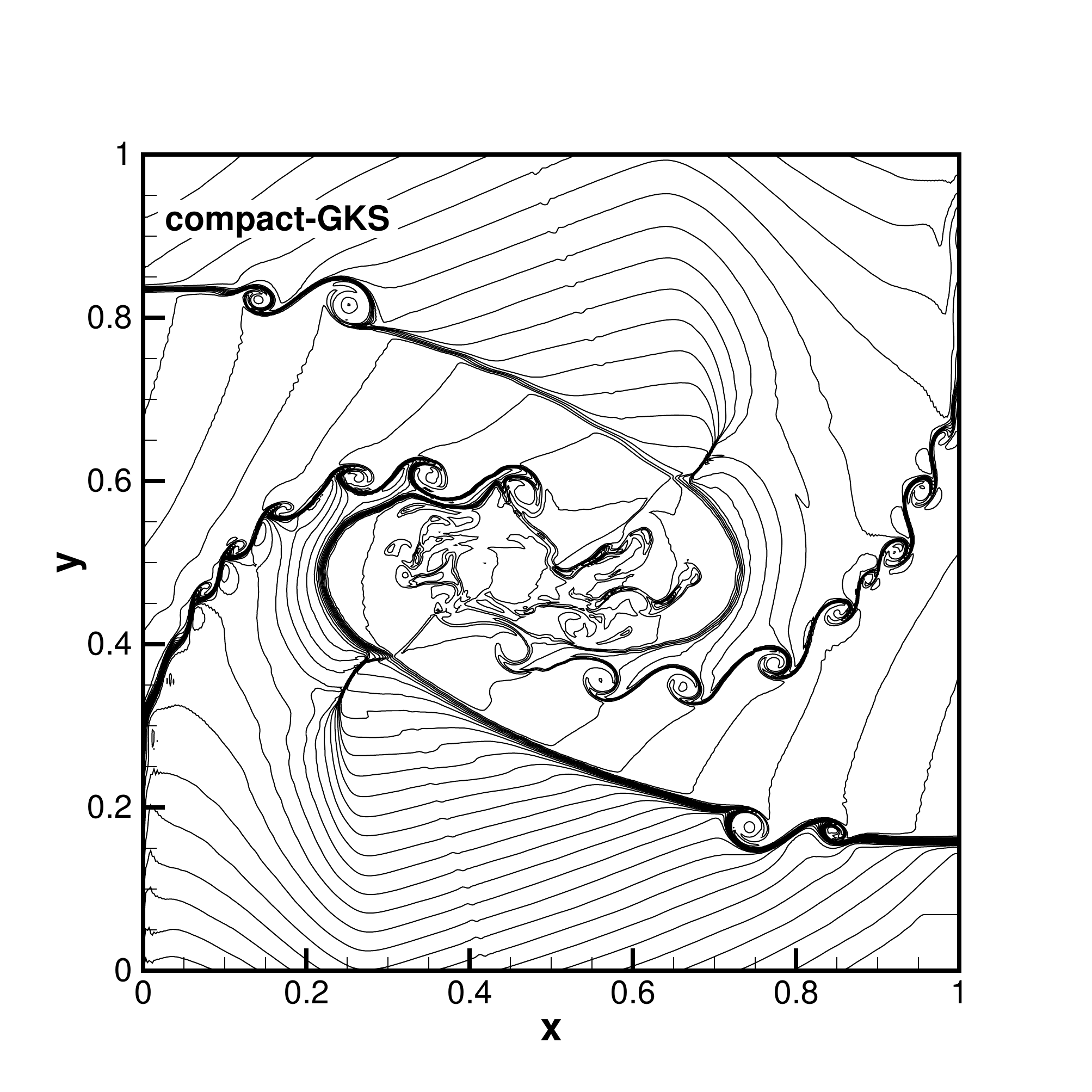}
\caption{\label{rm2d-shear-500-2} Two-dimensional Riemann problem:
the density distribution of four contact discontinuities-interaction
at $t=0.8$.}
\end{figure}

The initial conditions for the second 2-D Riemann problem are
\begin{equation*}
(\rho,U,V,p)=\left\{\begin{aligned}
         &(1, 0.1, 0.1, 1), &x>0.5,y>0.5,\\
         &(0.5197,-0.6259, 0.1, 0.4), &x<0.5,y>0.5,\\
         &(0.8, 0.1, 0.1, 0.4), &x<0.5,y<0.5,\\
         &(0.5197,0.1,-0.6259, 0.4), &x>0.5,y<0.5.
                          \end{aligned} \right.
                          \end{equation*}
This case is to simulate the shear instabilities among four initial
contact discontinuities. The density distributions calculated by
compact and non-compact GKS with HWENO and WENO reconstructions are presented at $t=0.8$ in Fig.\ref{rm2d-shear-500-2}.
The results indicate that the
current HWENO compact GKS resolves the Kelvin-Helmholtz instabilities better.

\begin{figure}[!h]
\centering
\includegraphics[width=0.8\textwidth]{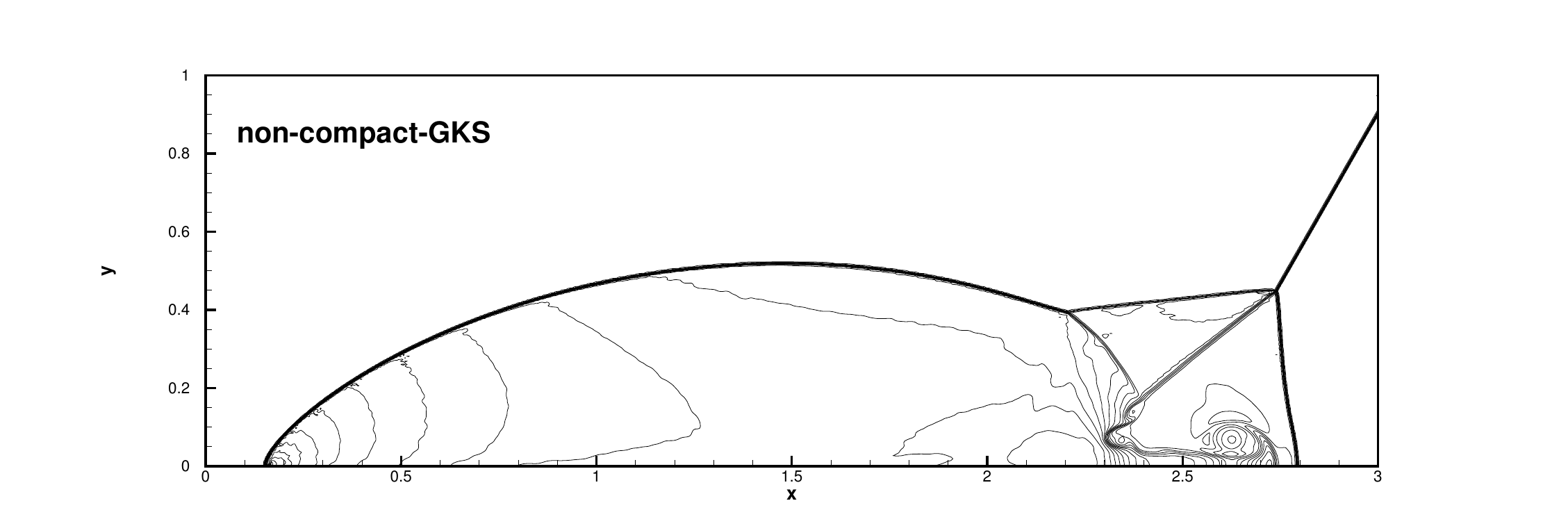}
\includegraphics[width=0.8\textwidth]{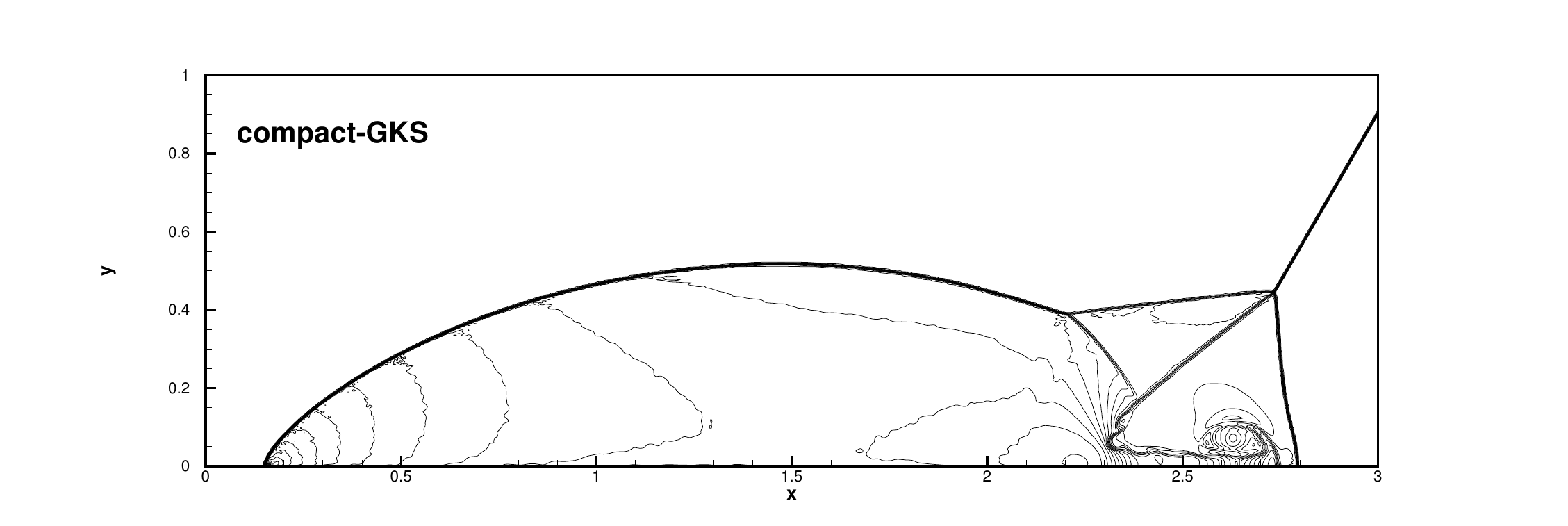}\\
\includegraphics[width=0.45\textwidth]{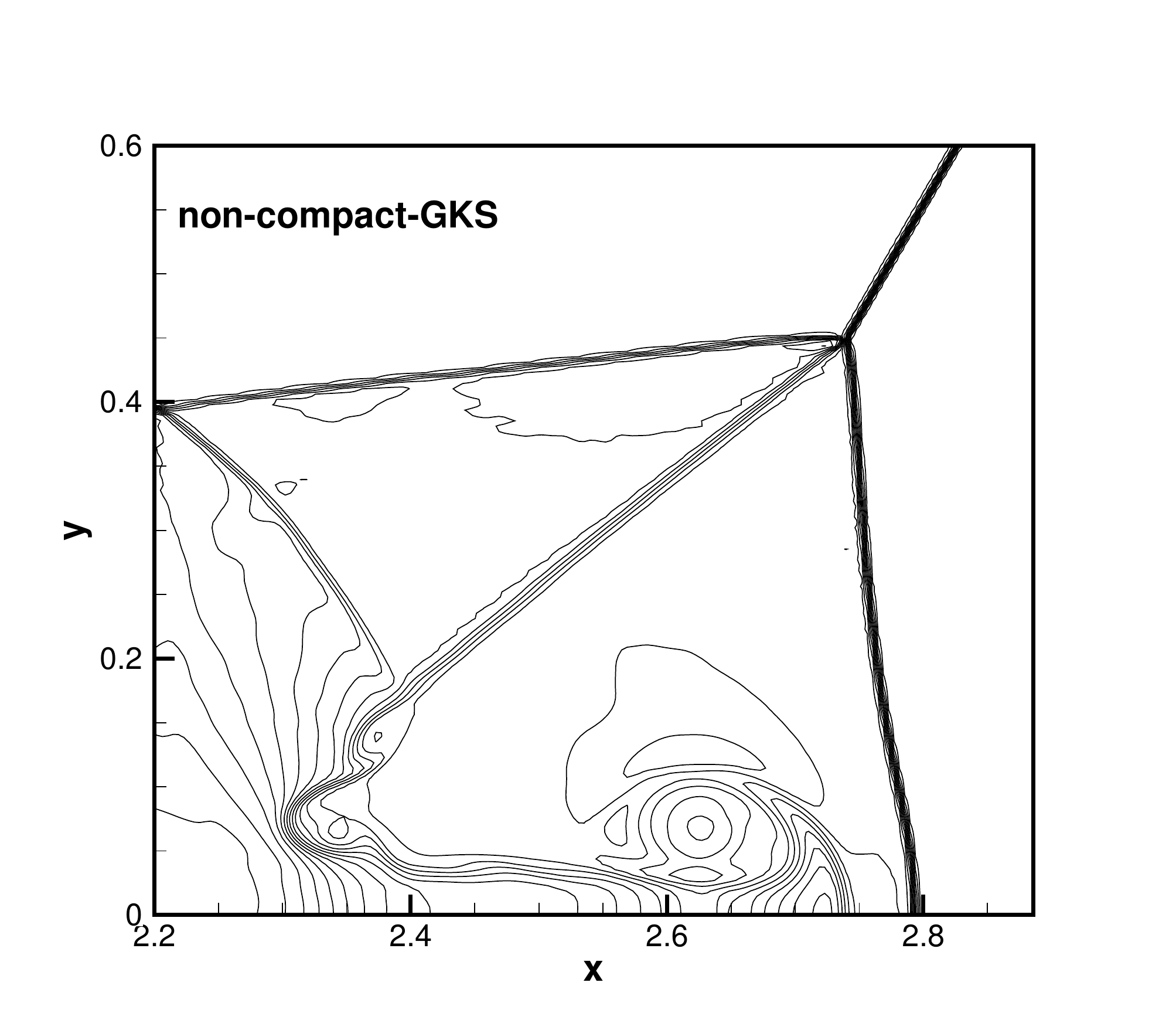}
\includegraphics[width=0.45\textwidth]{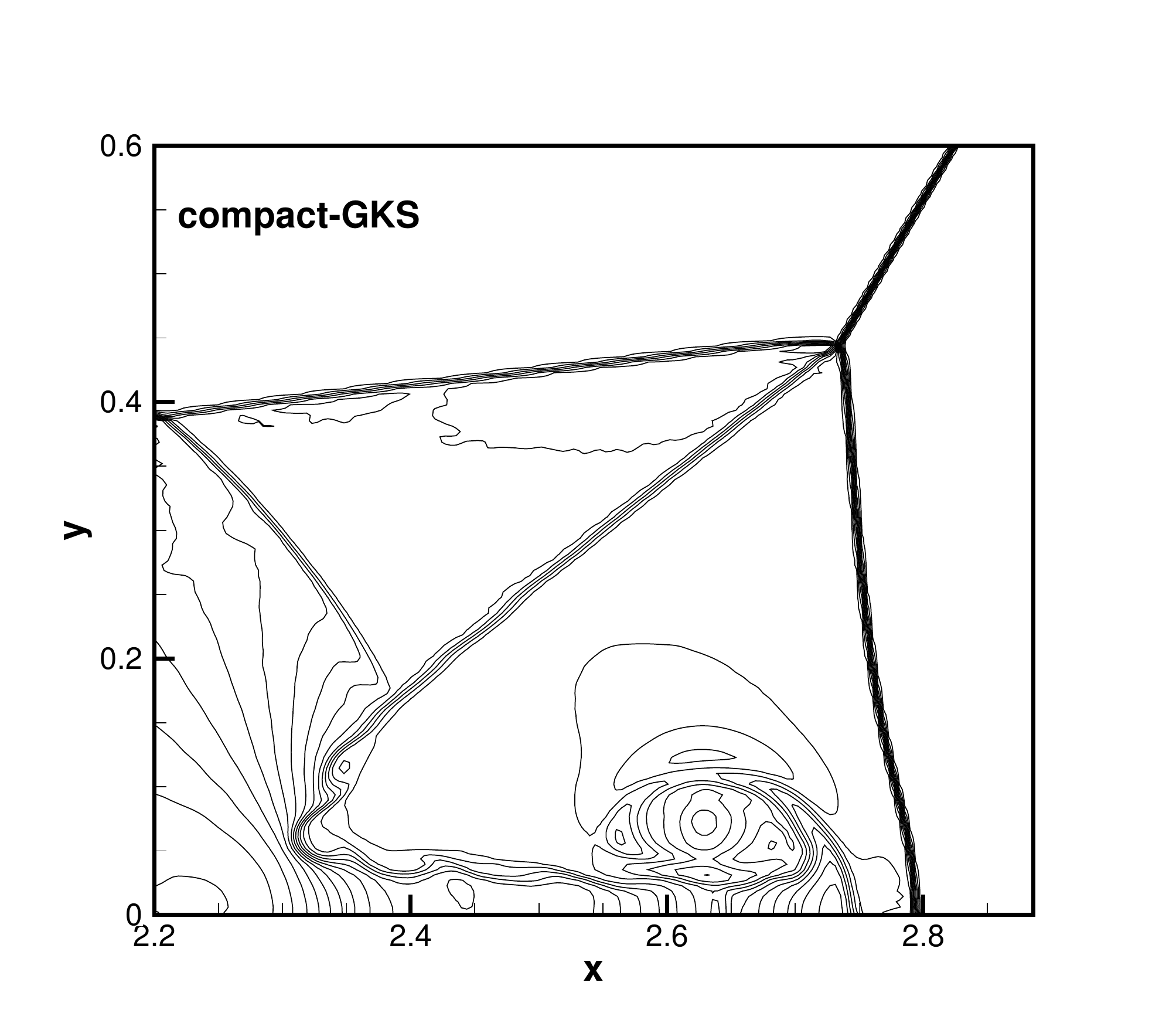}
\caption{\label{double-mach-1} Double Mach reflection: density
contours from compact and non-compact GKS with HWENO and WENO reconstructions and $960\times240$
mesh points.}
\end{figure}

\begin{figure}[!h]
\centering
\includegraphics[width=0.8\textwidth]{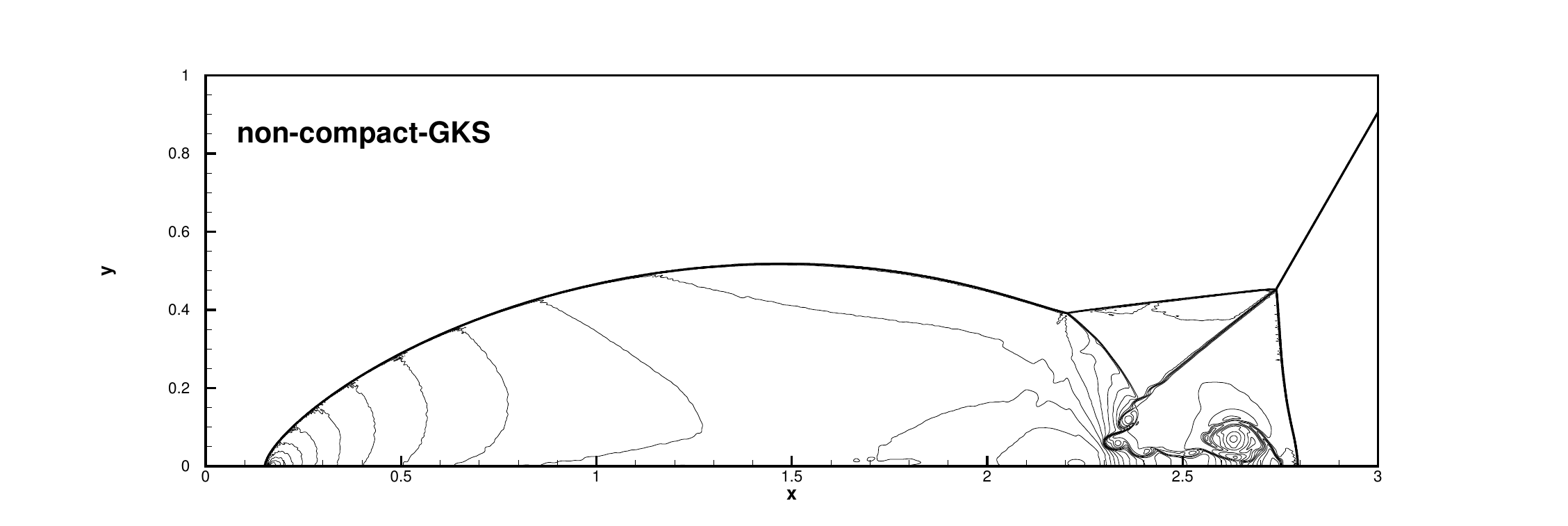}
\includegraphics[width=0.8\textwidth]{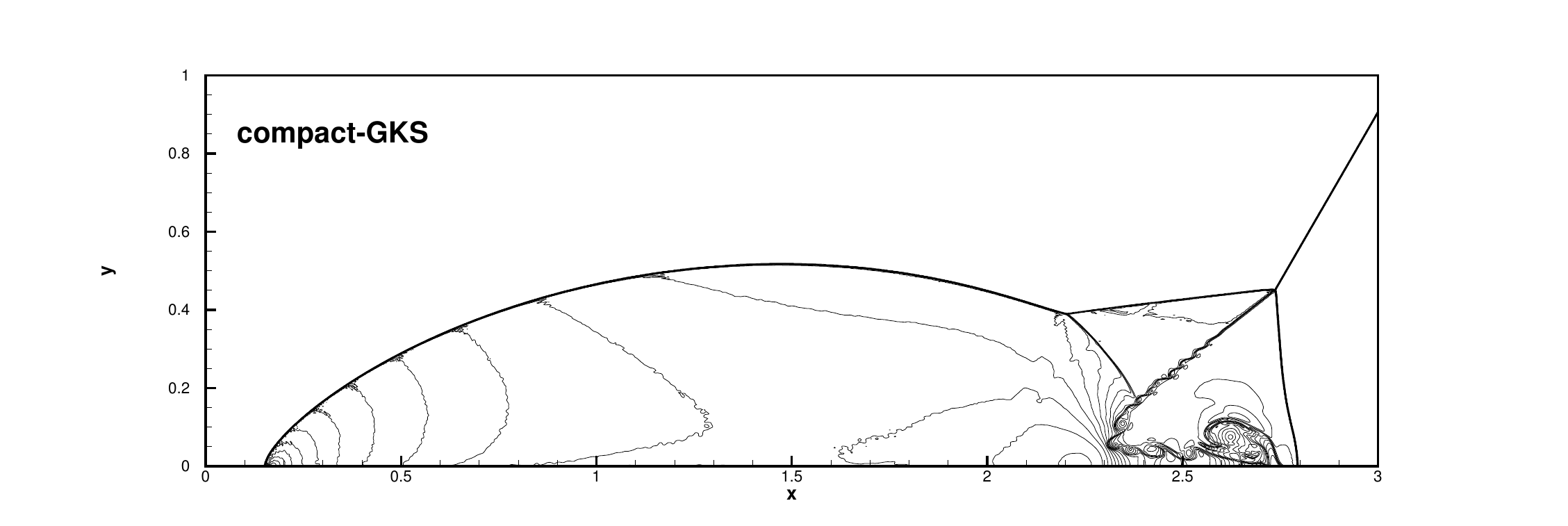}\\
\includegraphics[width=0.48\textwidth]{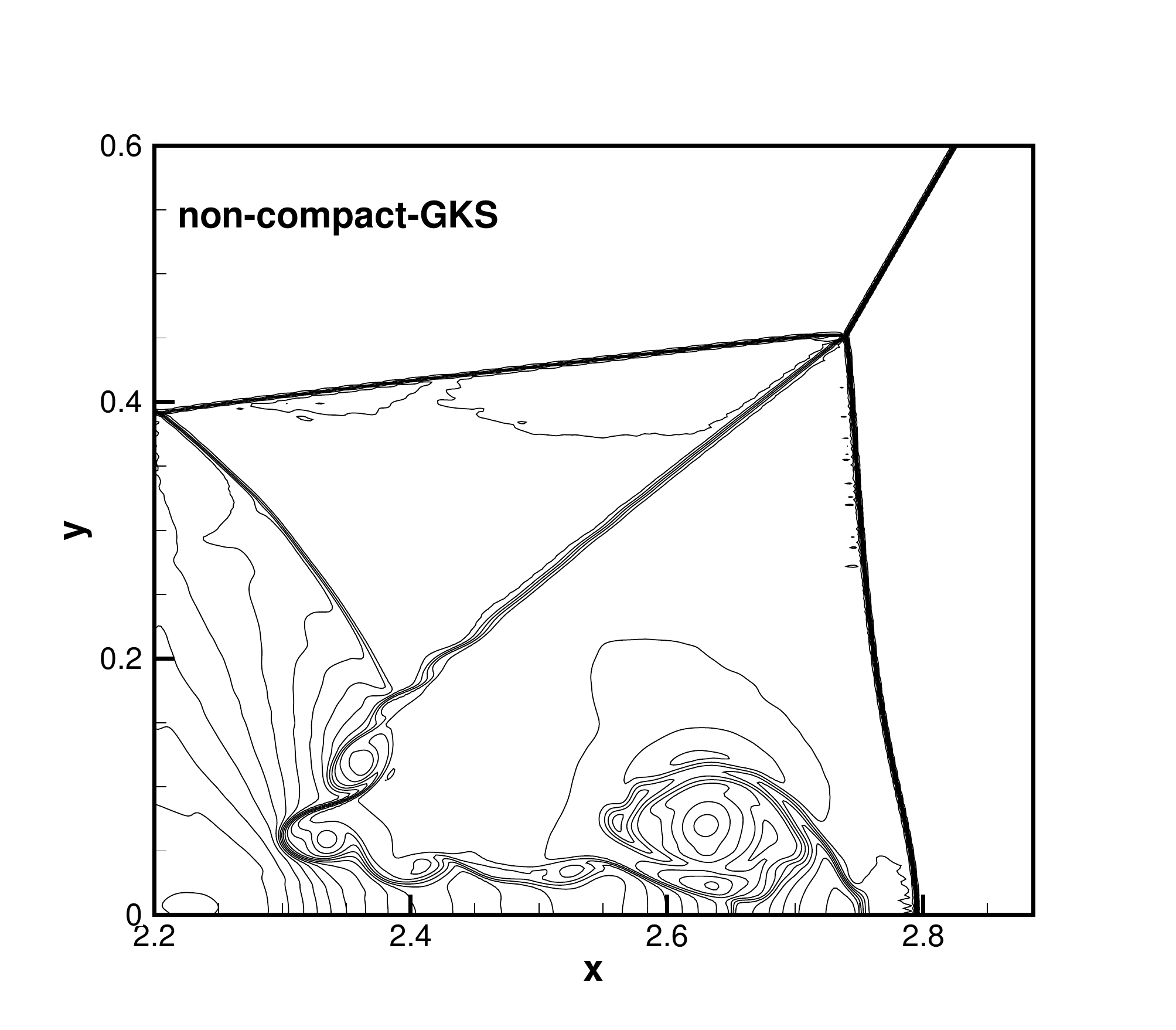}
\includegraphics[width=0.48\textwidth]{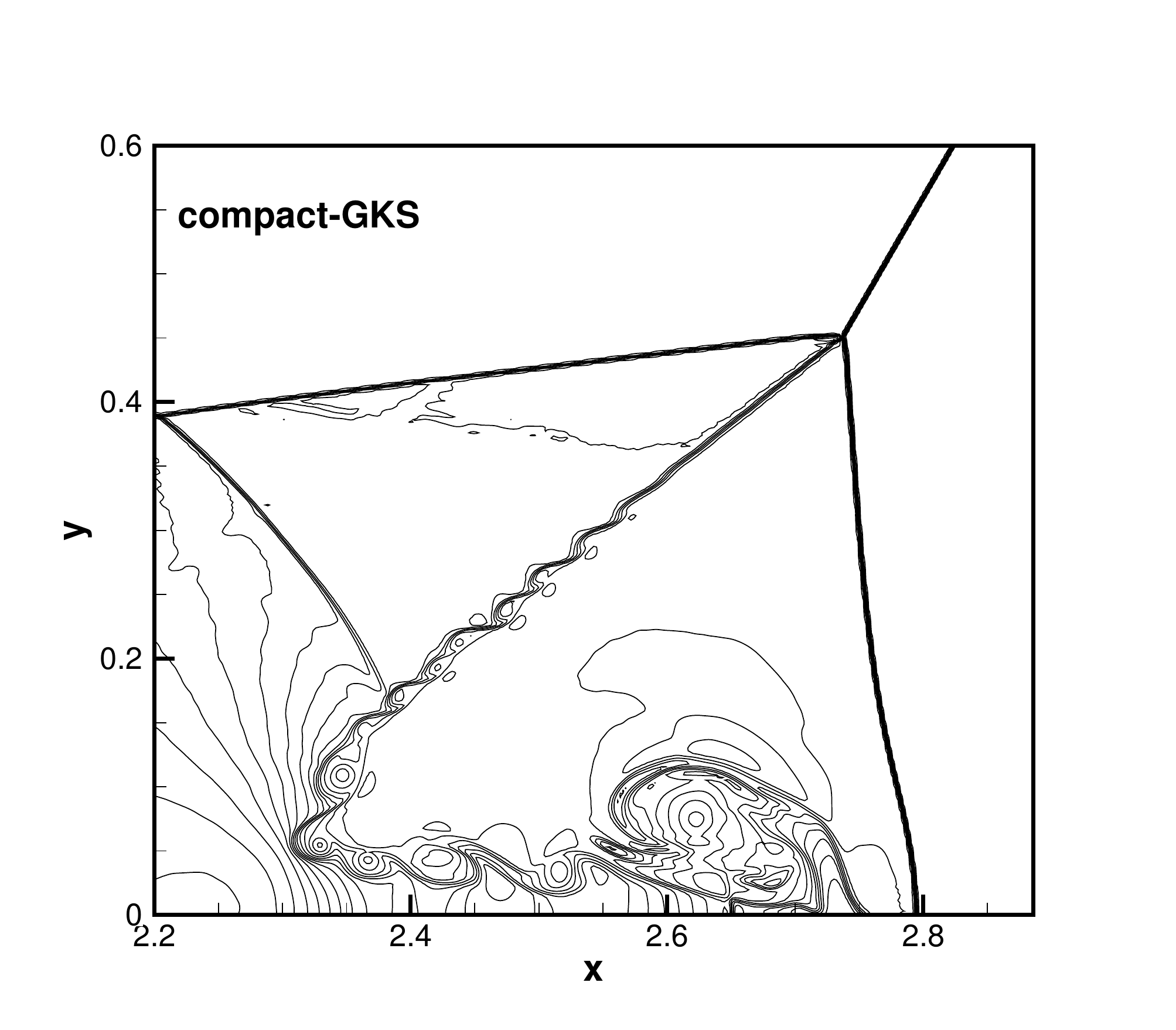}
\caption{\label{double-mach-2}  Double Mach reflection: local
enlargement of density contours from compact and non-compact GKS with HWENO and WENO reconstructions
and $1920\times480$ mesh points.}
\end{figure}

\subsection{Double Mach reflection problem}
This problem was extensively studied by Woodward and Colella
\cite{Woodward-Colella} for the inviscid flow. The computational
domain is $[0,4]\times[0,1]$, and a solid wall lies at the bottom of
the computational domain starting from $x =1/6$. Initially a
right-moving Mach $10$ shock is positioned at $(x,y)=(1/6, 0)$, and
makes a $60^\circ$ angle with the x-axis. The initial pre-shock and
post-shock conditions are
\begin{align*}
(\rho, U, V, p)&=(8, 4.125\sqrt{3}, -4.125,
116.5),\\
(\rho, U, V, p)&=(1.4, 0, 0, 1).
\end{align*}
The reflecting boundary condition is used at the wall, while for the
rest of bottom boundary, the exact post-shock condition is imposed.
At the top boundary, the flow variables are set to follow  the
motion of the Mach $10$ shock. The density distributions and
local enlargement with $960\times240$ and $1920\times480$ uniform
mesh points at $t=0.2$ with HWENO reconstructions are shown in
Fig.\ref{double-mach-1} and Fig.\ref{double-mach-2}. The robustness
of the compact GKS is validated, and the flow structure
around the slip line from the triple Mach point is resolved better by the compact scheme.

\begin{table}[!h]
\begin{center}
\def\temptablewidth{1.0\textwidth}
{\rule{\temptablewidth}{0.5pt}}
\begin{tabular*}{\temptablewidth}{@{\extracolsep{\fill}}ccccc}
Scheme & AUSMPW+ \cite{Case-Kim} & M-AUSMPW+ \cite{Case-Kim} & WENO-GKS & HWENO-GKS  \\
\hline
Height  & 0.163   &  0.168    & 0.171  &  0.173     \\
\end{tabular*}
{\rule{\temptablewidth}{0.5pt}}
\end{center}
\vspace{-4mm} \caption{\label{height}  Viscous shock tube problem:
comparison of the primary vortex heights among different schemes
with $500\times250$ uniform mesh points for $Re=200$ case.}
\end{table}

\subsection{Viscous shock tube problem}
This problem was introduced to test the performances of different
schemes for viscous flows \cite{Case-Daru}. In this case, an ideal
gas is at rest in a two-dimensional unit box $[0,1]\times[0,1]$. A
membrane located at $x=0.5$ separates two different states of the
gas and the dimensionless initial states are
\begin{equation*}
(\rho,U,p)=\left\{\begin{aligned}
&(120, 0, 120/\gamma), \ \ \ &  0<x<0.5,\\
&(1.2, 0, 1.2/\gamma),  & 0.5<x<1,
\end{aligned} \right.
\end{equation*}
where $\gamma=1.4$ and Prandtl number $Pr=0.73$.

\begin{figure}[!h]
\centering
\includegraphics[width=0.65\textwidth]{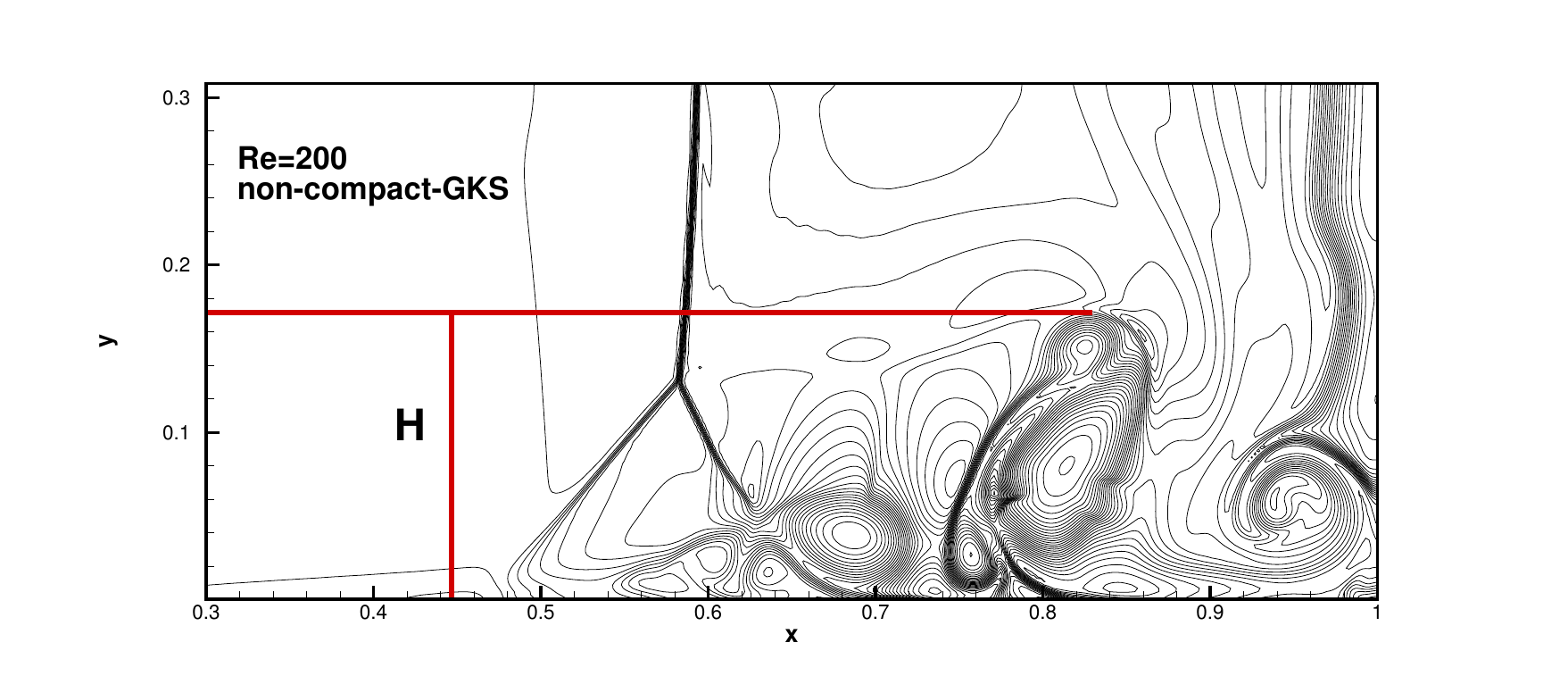}
\includegraphics[width=0.65\textwidth]{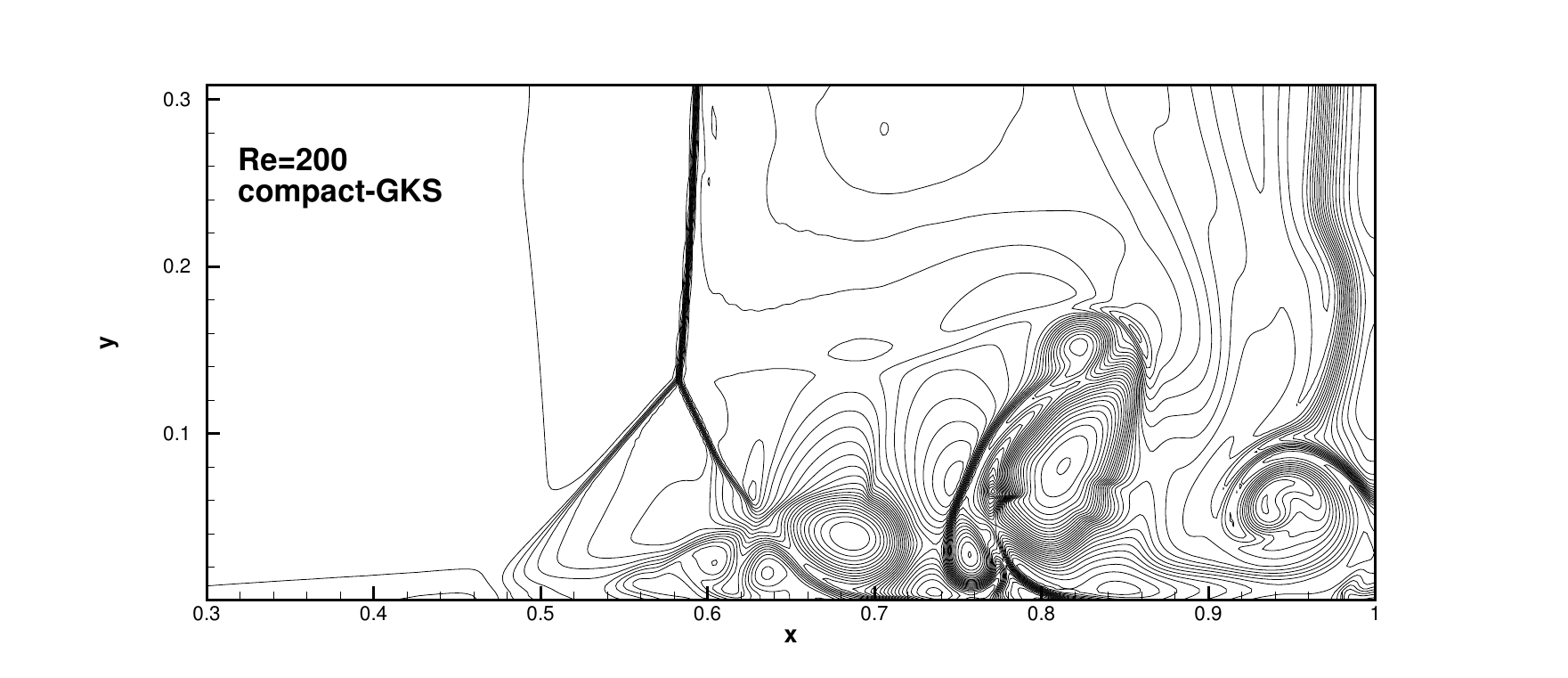}
\caption{\label{vistube-re200-1} Viscous shock tube problem: density
contours with $500\times250$ uniform mesh points at $t=1$ for
$Re=200$ case.}
\centering
\includegraphics[width=0.6\textwidth]{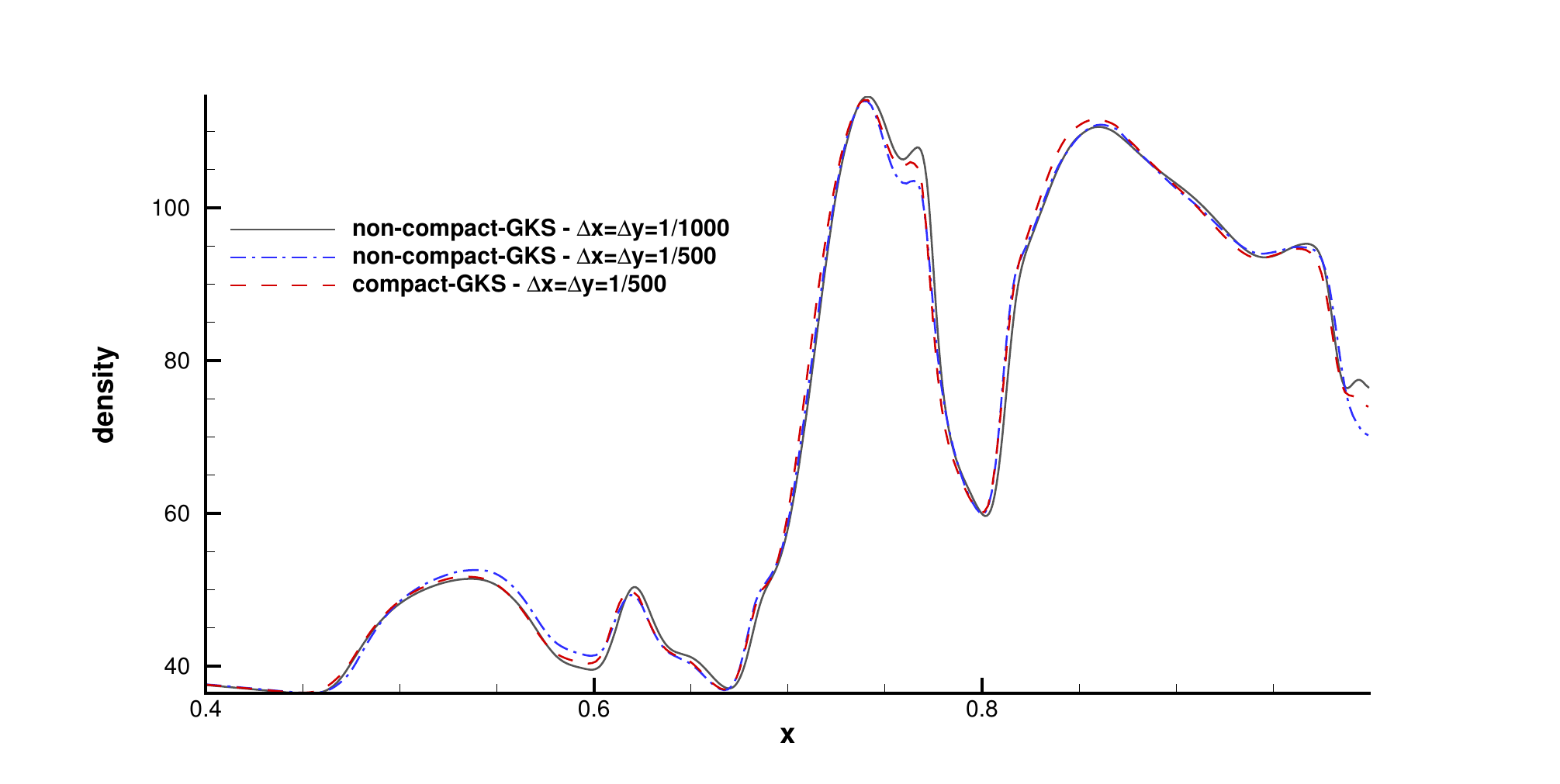}
\caption{\label{vistube-re200-2} Viscous shock tube problem: density
profiles along the lower wall at $t=1$ for $Re=200$ case. }
\end{figure}

\begin{figure}[!h]
	\centering
	\includegraphics[width=0.65\textwidth]{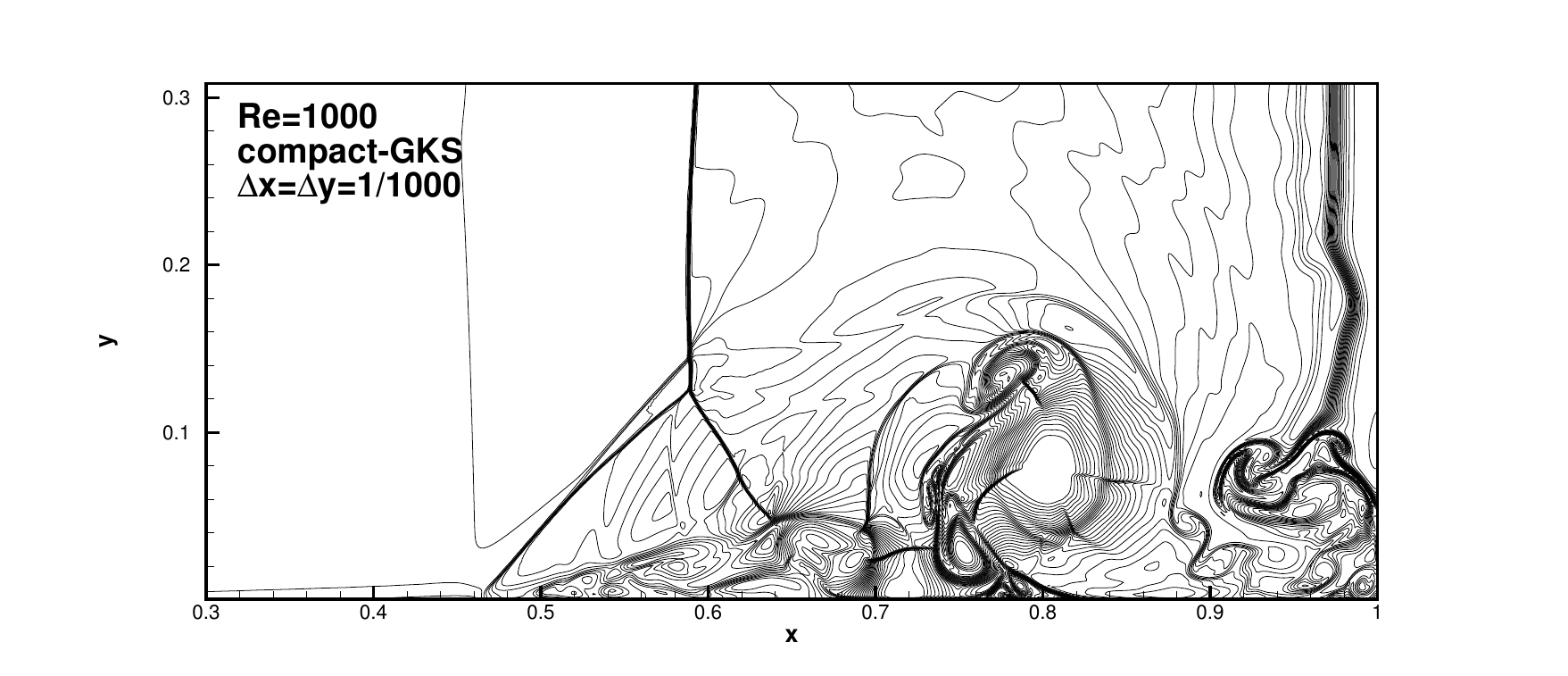}
	\caption{\label{vistube-re1000-1} Viscous shock tube problem: density
		contours with $1000\times500$ uniform mesh points at $t=1$ for
		$Re=1000$ case.}
\end{figure}

\begin{figure}[!h]
	\centering
	\includegraphics[width=0.65\textwidth]{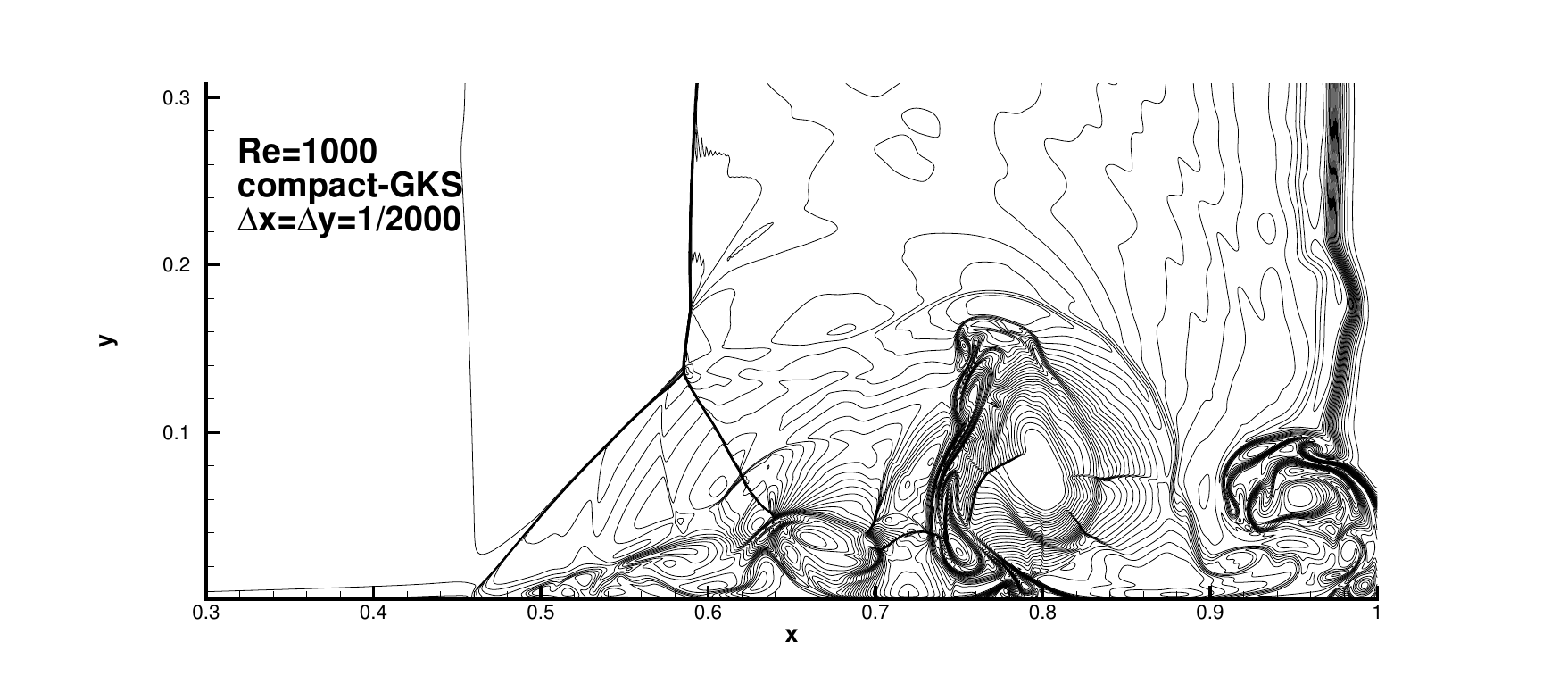}
	\caption{\label{vistube-re2000-1} Viscous shock tube problem: density
		contours with $2000\times1000$ uniform mesh points at $t=1$ for
		$Re=1000$ case.}
\end{figure}

\begin{figure}[!h]
	\centering
	\includegraphics[width=0.65\textwidth]{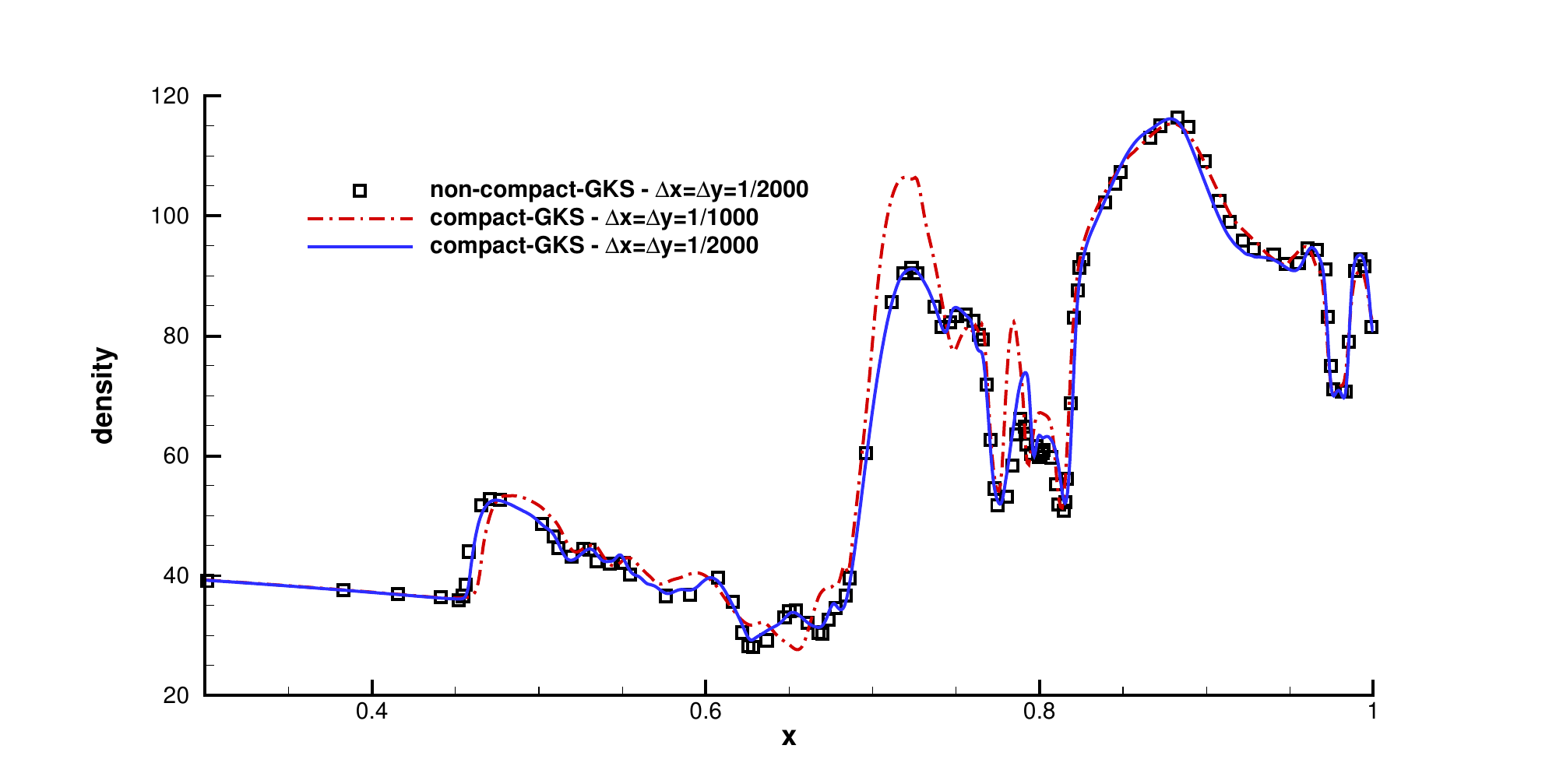}
	\caption{\label{vistube-re1000-2}  Viscous shock tube problem: density
		profiles along the lower surface at $t=1$ for $Re=1000$ case.}
\end{figure}

The membrane is
removed at time zero and wave interaction occurs. A shock wave,
followed by a contact discontinuity, moves to the right with Mach
number $Ma=2.37$ and reflects at the right end wall. After the
reflection, it interacts with the contact discontinuity. The contact
discontinuity and shock wave interact with the horizontal wall and
create a thin boundary layer during their propagation. The solution
will develop complex two-dimensional shock/shear/boundary-layer
interactions. This case is tested in the computational domain $[0,
1]\times[0, 0.5]$, a symmetric boundary condition is used on the top
boundary $x\in[0, 1], y=0.5$. Non-slip boundary condition, and
adiabatic condition for temperature are imposed at solid wall.
Firstly, the Reynolds number $Re=200$ case is tested.
For this case with $Re=200$, the density distributions  with
$500\times250$ uniform mesh points at $t=1.0$ from non-compact and compact GKS with HWENO and WENO
reconstructions are shown in Fig.\ref{vistube-re200-1}. The density
profiles along the lower wall for this case are presented in
Fig.\ref{vistube-re200-2}. As a comparison, the results from WENO
reconstruction with $1000\times500$ uniform mesh points is given as
well, which agrees well with the density profiles provided by compact GKS with HWENO
method and $500\times250$ mesh points.
As shown in Table.\ref{height}, the
height of primary vortex predicted by the current compact scheme agrees well
with the reference data \cite{Case-Kim}.

Secondly, the $Re=1000$ case is computed with different girds.
For the case with $1000 \times 500$ coarse mesh points vortex shedding could be observed clearly at the wedge-shaped area defined in \cite{Case-Zhou}, seen in Fig.\ref{vistube-re1000-1}.
Also, the density distribution along the wall at $t=1.0$ is plotted in Fig.\ref{vistube-re1000-2}. In comparison with the reference result of two stage fourth order GKS \cite{GKS-high3}, both the overall density contours, seen in Fig.\ref{vistube-re2000-1} and density distribution along the wall agree well with traditional non-compact WENO GKS.
 \begin{figure}[!h]
 	\centering

    \includegraphics[width=0.44\textwidth]{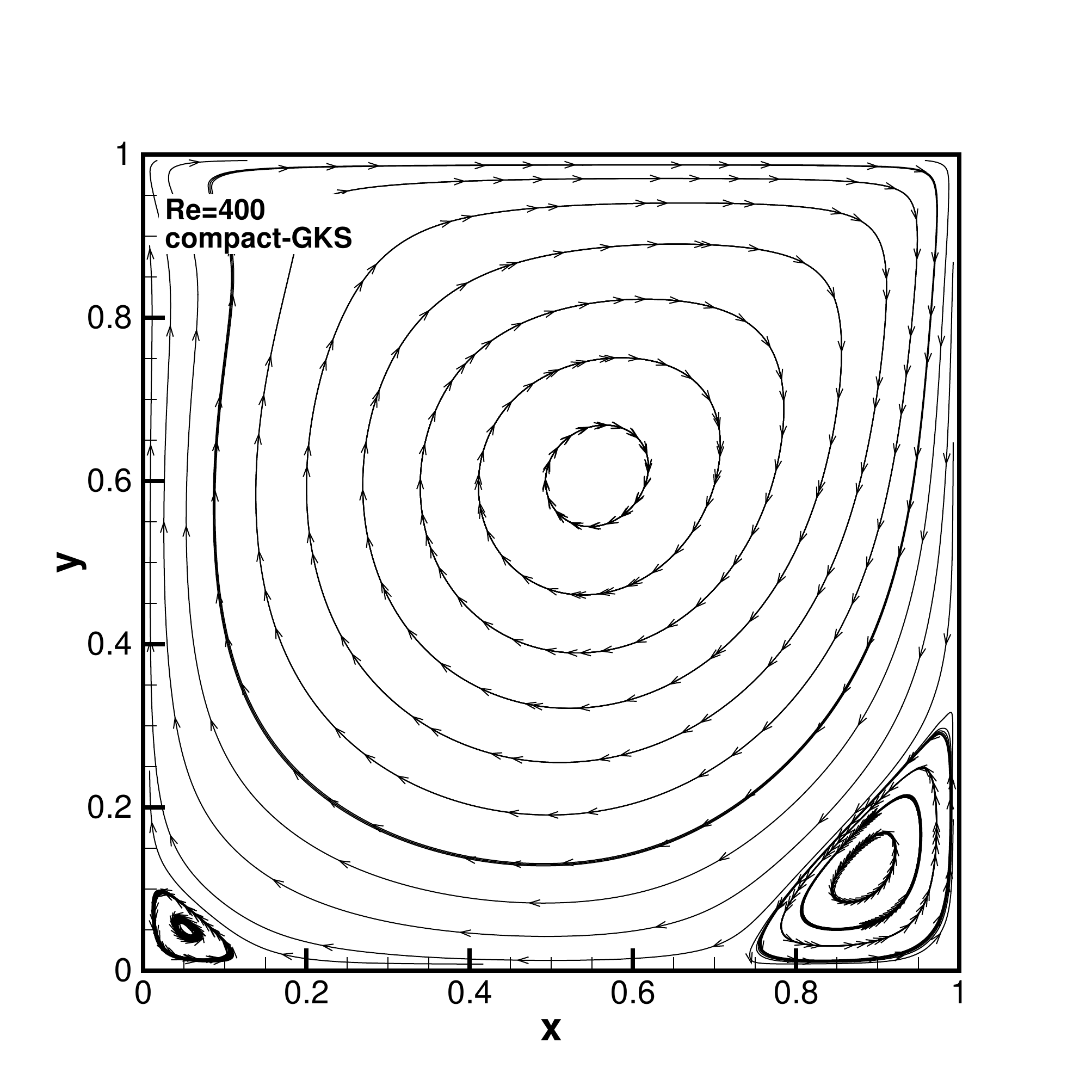}
 	\includegraphics[width=0.44\textwidth]{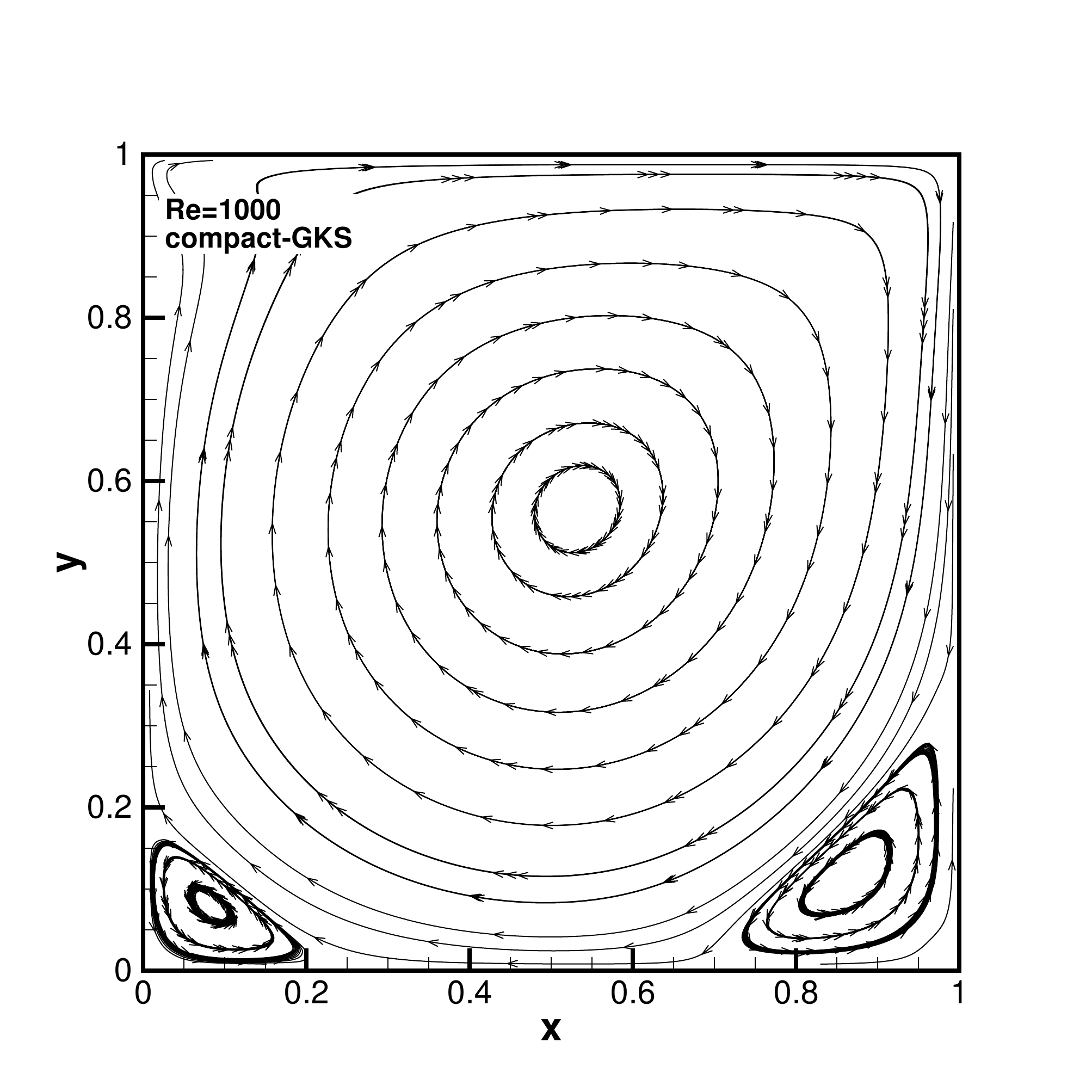}
 	\caption{\label{cavity-1} Lid-driven cavity flow: streamlines with $65\times65$ uniform mesh points
 	 for $Re=400$ and $Re=1000$ by compact HWENO-GKS.}
 	\centering
 	\includegraphics[width=0.44\textwidth]{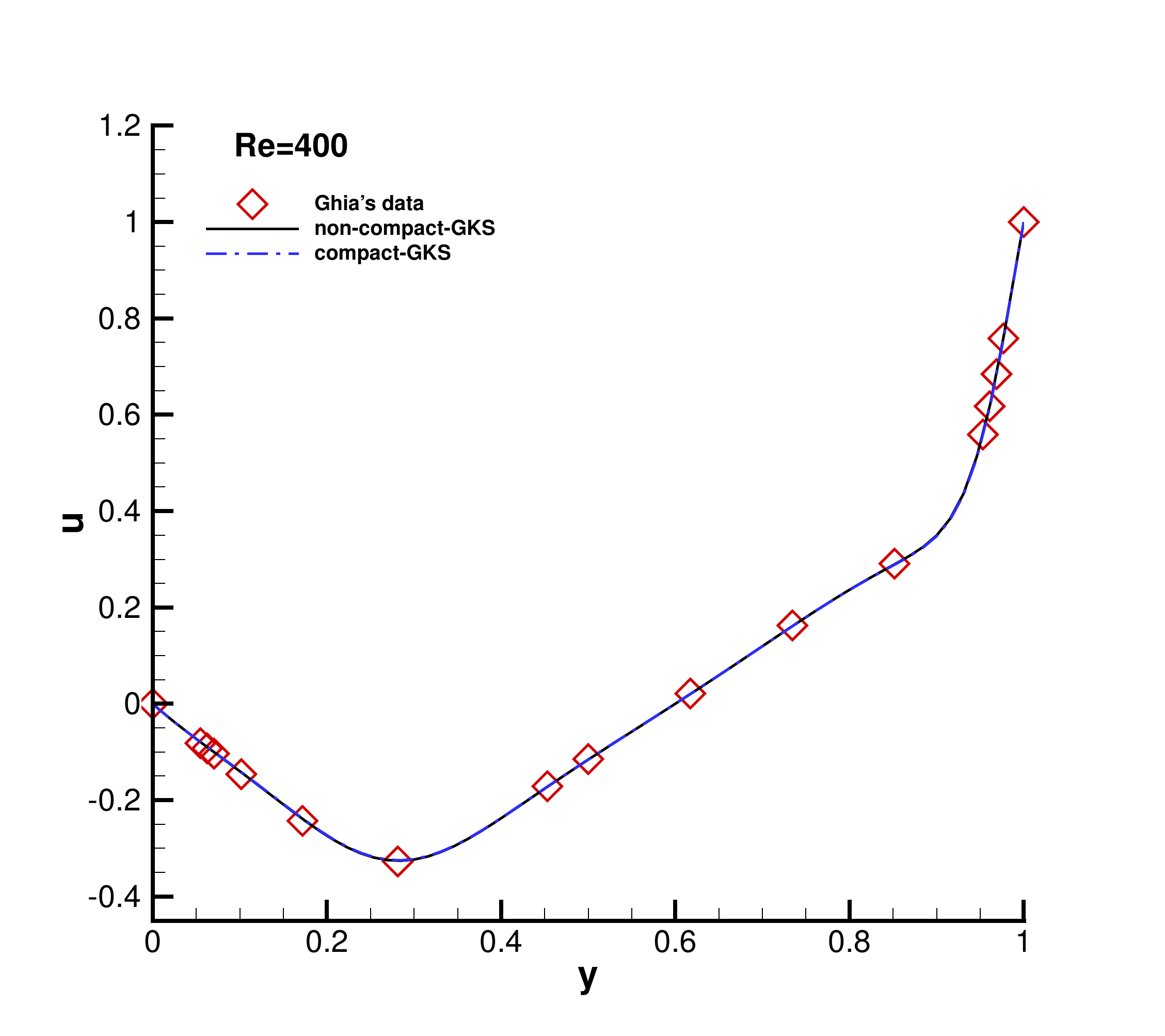}
 	\includegraphics[width=0.44\textwidth]{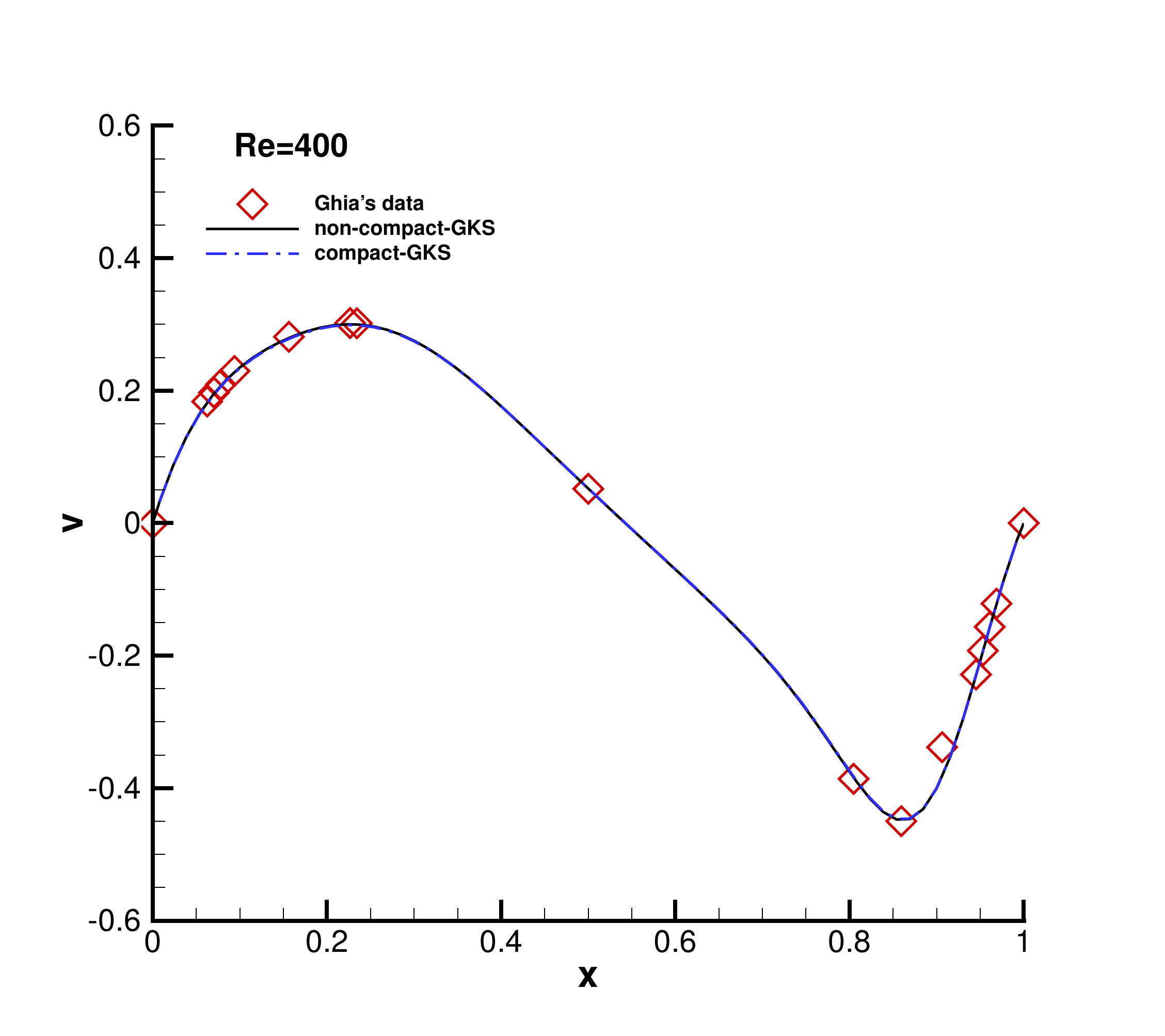}
 	 	\includegraphics[width=0.44\textwidth]{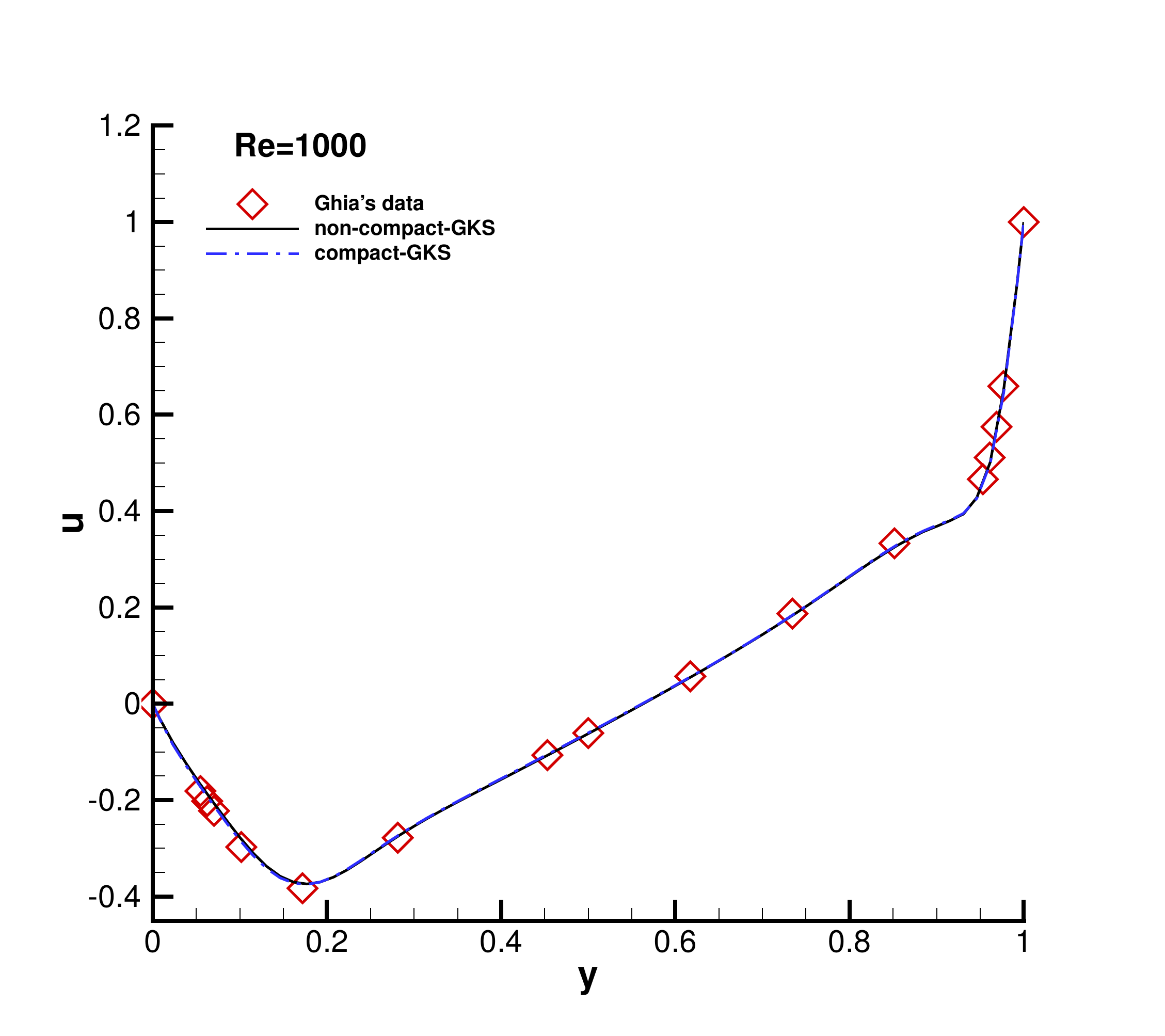}
 	\includegraphics[width=0.44\textwidth]{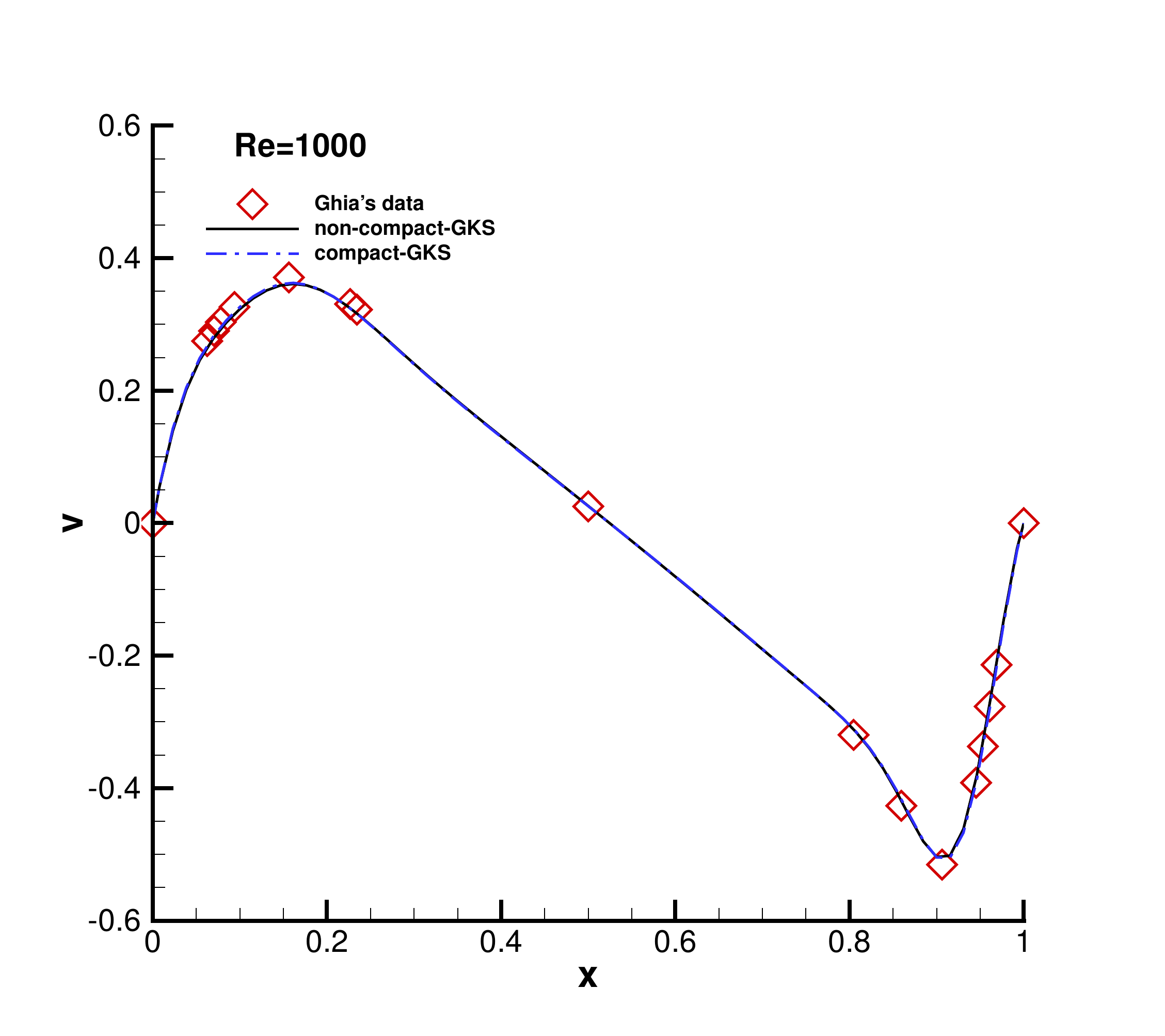}
 	\caption{\label{cavity-2} Lid-driven cavity flow: Comparisons between non-compact WENO-GKS and compact HWENO-GKS by U-velocity along vertical centerline line and V-velocity along horizontal center-line with $65\times65$ uniform mesh points, at $Re=400$ and $1000$. }
 \end{figure}

\subsection{Lid-driven cavity flow}
In order to further test the scheme in the capturing of viscous flow solution,
the lid-driven cavity problem is one of the most important
benchmarks for validating incompressible  Navier-Stokes
flow solvers. The fluid is bounded by a unit square and is driven by a
uniform translation of the top boundary. In this case, the flow is
simulated with Mach number $Ma=0.15$ and all boundaries are
isothermal and nonslip. The computational domain $[0, 1]\times[0,1]$ is covered with $65\times65$ mesh points.
Numerical simulations are conducted for two different Reynolds numbers, i.e.,
$Re=400$ and $1000$. The streamlines in Fig.\ref{cavity-1}, the
$U$-velocities along the center vertical line, and $V$-velocities along
the center horizontal line, are shown in Fig.\ref{cavity-2}. The
benchmark data \cite{Case-Ghia} for $Re=400 $ and $ 1000$ are also presented,
and  the simulation results match well with these benchmark data.
The cavity case fully validates the higher-order accuracy of the compact GKS. With $65\times 65$ mesh points,
second-order schemes cannot get such accurate solutions.

\section{Conclusion}
In this paper, a fourth-order compact gas-kinetic scheme based on
Hermite WENO reconstruction is presented.
The construction of such a compact higher-order scheme is solely due to the use of the updated cell
interface values, from which the averaged slopes inside each cell can be obtained.
Therefore, the HWENO reconstruction can be naturally implemented here.
There are similarity and differences between the current GKS and the compact 4th-order DG method.
Both schemes have the same order accuracy and use the same HWENO reconstruction with the same compactness of the stencil.
The main difference  between these two methods is that
for the DG method  the slope at time step $t^{n+1}$ is obtained through time evolution of the slope directly.
However, for the GKS the cell interface values  are  evolved first, then the slope at $t^{n+1}$ is derived using Gauss's law through the interface
values.
More specifically, in the DG method the cell averaged values through fluxes and their slopes are evolved separately with
individual discretized governing equations, while in the GKS the updates of both cell averaged values and their slopes are coming from the
same interface time-dependent gas distribution function, which is an analytically evolution solution of the kinetic relaxation model.
The DG is based on the weak formulation with the involvement of test function, the GKS is based on the strong solution, which is unique from the
kinetic model equation and the initial reconstruction.
As a result, for  the 4th-order accuracy, the DG uses the Runge-Kutta time stepping scheme with four stages within each time step and has a CFL number limitation of $0.11$ from the stability consideration. For the same order GKS, only two stages are involved and the CFL number for the time step can be on
the order of $0.5$. Even with the expensive GKS flux function targeting on the Navier-Stokes solutions,
the computational efficiency  of the GKS is much higher than that of the DG method for the Euler equations alone.
Based on the test cases in this paper and many others not presented here,
the 4th-order compact GKS has the same robustness as the 2nd-order
shock capturing schemes, where there are no trouble cells and any other special limiting process involved in the GKS calculations.
For the high speed compressible flow it is still an active research subject in the DG formulation to improve its robustness
to very complicated flow interactions, such as
the computation of viscous shock tube case at Reynolds number $1000$,
and there is no any clear direction for its further improvement.
Many numerical examples are
presented to validate the higher-order compact GKS. The current paper only presents the compact scheme on structured rectangular mesh.
Following the approach of the 3rd-order compact GKS on unstructured mesh \cite{pan-xu1},
the current 4th-order compact GKS is being extended there as well.

The present research clearly indicates that the dynamics of the 1st-order Riemann solver is not enough for the construction of truly higher-order
compact scheme. To keep the compactness is necessary for any scheme with correct physical modeling because the evolution of gas dynamics in any scale from the kinetic particle transport to the hydrodynamic wave propagation does
only involve neighboring flow field with limited propagating speed \cite{xu-liu}, where the CFL condition is not only a stability requirement, but also quantifies the relative physical domain of dependence. Ideally, the numerical domain of dependence should be the same as the physical domain of
dependence, rather than the large disparity between them in the current existing schemes, such as the very large numerical domain of
 dependence (large stencils) in the WENO approach and the severely confined CFL number in the DG method.
Due to the local high-order gas evolution model in GKS, both numerical and physical domains of dependence are getting closer in the current compact
GKS for the Euler and NS solutions. The unified gas kinetic scheme (UGKS) makes these two domains of dependence even closer  than that of GKS \cite{GKS-Xu1,xu-liu},
such as in the low Reynolds number viscous flow computations, where the UGKS can use a much larger CFL number than that of GKS.

\section*{Acknowledgement}
The authors would like to than Prof. J.Q. Li and J.X. Qiu for helpful discussion.
The current research was supported by Hong Kong research grant council (16206617,16207715,16211014)
and  NSFC (91530319,11772281).

\section*{Appendix 1: HWENO Reconstruction at Gaussian points}
For the interface values reconstruction, starting from the same stencils as in \cite{HWENO1}, the
pointwise values for three sub-stencils at Gaussian point
$x_{i+1/2\sqrt{3}}$ are
\begin{align*}
p_{0}(x_{i+1/2\sqrt{3}})&=-\frac{1}{\sqrt{3}}W_{i-1}+\frac{3+\sqrt{3}}{\sqrt{3}}W_i-\frac{\Delta x}{2 \sqrt{3}}(W_x)_{i-1},\\
p_{1}(x_{i+1/2\sqrt{3}})&=\frac{3-\sqrt{3}}{\sqrt{3}}W_i+\frac{1}{\sqrt{3}}W_{i+1}-\frac{\Delta x}{2 \sqrt{3}}(W_x)_{i+1},\\
p_{2}(x_{i+1/2\sqrt{3}})&=-\frac{1}{4\sqrt{3}}W_{i-1}+W_i+\frac{1}{4
\sqrt{3}}W_{i+1}.
\end{align*}
The pointwise value for the large stencil at the Gaussian point
$x_{i+1/2\sqrt{3}}$ is
\begin{align*}
q(x_{i+1/2\sqrt{3}})=\frac{1}{720}[&(2-95 \sqrt{3})W_{i-1} +(2+95 \sqrt{3})W_{i+1}+716W_i\\
+\Delta x&(1-35 \sqrt{3})(W_x)_{i-1} -\Delta x
(1+35\sqrt{3})(W_x)_{i+1})].
\end{align*}
In order to satisfy the following equations
\begin{align*}
q(x_{i+1/2\sqrt{3}})=\sum_{k=0}^{2}\hat{\gamma}_{k}p_{k}(x_{i+1/2\sqrt{3}}),
\end{align*}
the linear weights become $\displaystyle \hat{\gamma}_{0}=\frac{105-
\sqrt{3}}{360}, \hat{\gamma}_{1}=\frac{105+ \sqrt{3}}{360},
\hat{\gamma}_{2}=\frac{5}{12}$.  The smoothness indicators and
non-linear weights take the identical forms as that
in \cite{HWENO1}. The pointwise values and linear weights at
another Gaussian point $x_{i-1/2\sqrt{3}}$ can be obtained similarly using
symmetrical property.

\section*{Appendix 2: Reconstruction for slope at Gaussian points}
The slope at Gaussian points is reconstructed as follows.
Denote that $P_{i-1/2,j_l}, l=0,1$ as the Gaussian quadrature points
of the interface $(i-1/2,j)$. For simplicity, the suffix $i-1/2$ is omitted next without confusion.
Two sub-stencils are defined as
\begin{align*}
S_0=\{P_{(j-1)_2}, P_{j_1}, P_{j_2}\}, ~~S_1=\{P_{j_1}, P_{j_2}, P_{(j+1)_1}\},
\end{align*}
where $j_1=j-1/2\sqrt{3}, j_2=j+1/2\sqrt{3}$. A quadratic polynomial
$p_i(y)$ can be constructed corresponding to $S_i, i=0,1$. For the
large stencil
\begin{align*}
\mathbb{T}=\{S_0, S_1\},
\end{align*}
a cubic polynomial $q(y)$ can be constructed as well. The first-order
derivative of $q(y)$ can be written as a linear combination of all
first-order derivatives of $p_i(y)$ corresponding to stencils $S_i$ at
Gaussian point $y_{j+1/2\sqrt{3}}$
\begin{align*}
\frac{\partial q}{\partial y}(y_{j+1/2\sqrt{3}})=\sum_{k=0}^{1}
\check { \gamma} _{k}\Big(\frac{\partial p}{\partial
y}\Big)_{k}(y_{j+1/2\sqrt{3}}),
\end{align*}
where the linear weights takes $\displaystyle
\check{\gamma}_{0}=\frac{5- \sqrt{3}}{11},
\check{\gamma}_{1}=\frac{6+ \sqrt{3}}{11}$, and
\begin{align*}
\Big( \frac{\partial p}{\partial
y}\Big)_{0}(y_{j+1/2\sqrt{3}})&=\frac{1}{(\sqrt{3}-1)\Delta y}
(W_{(j-1)_2}-3W_{j_1}+2W_{j_2}),\\
\Big(\frac{\partial p}{\partial
y}\Big)_{1}(y_{j+1/2\sqrt{3}})&=\frac{1}{(\sqrt{3}-1)\Delta y}[(-2+
\sqrt{3})W_{j_1}+(3-2\sqrt{3})W_{j_2}+W_{(j+1)_1}].
\end{align*}
where  $W_{j_l}$ are the pointwise values at the Gaussian quadrature
points $P_{j_l}$. The smoothness indicators for
$\check{\beta}_k,~k=0,1$ are defined by
\begin{equation*}
\check{\beta_k}=\Delta
x^{3}\int_{y_{j-1/2}}^{y_{j+1/2}}\big(\frac{\text{d}^2}{\text{d}y^2}p_i(x)\big)^2\text{d}x,
\end{equation*}
and the detailed formulae are given as follows
\begin{align*}
\check{\beta_0}&=\frac{4}{(\sqrt{3}-1)^2}[\sqrt{3}W_{(j-1)_2}-3W_{j_1}+(3-\sqrt{3})W_{j_2}]^2,\\
\check{\beta_1}&=\frac{4}{(\sqrt{3}-1)^2}[(3-\sqrt{3})W_{j_1}-3W_{j_2}+\sqrt{3}W_{(j+1)_1}]^2.
\end{align*}
Finally, the non-normalized nonlinear weights $\overline{\check{\omega}}_{k}, k=0,1$ are
\begin{align*}
\overline{\check{\omega}}_{k}=\frac{\check{\gamma}_{k}}{(\check{\beta_k}+\epsilon)^2}.
\end{align*}
The pointwise values and linear weights at Gaussian point
$y_{j-1/2\sqrt{3}}$ can be obtained similarly using the symmetrical property.

\end{document}